\documentclass[aps,prx,twocolumn,superscriptaddress]{revtex4-1}
\usepackage{amssymb}
\usepackage{amsmath}
\usepackage{graphicx}
\usepackage[dvipsnames]{xcolor}
\usepackage[version=3]{mhchem}
\usepackage{footnote}
\usepackage{bm}
\usepackage{dsfont}
\usepackage{xr-hyper}
\usepackage{hyperref}
\usepackage{cleveref}

\crefname{equation}{Eq.}{Eqs.}
\Crefname{equation}{Equation}{Equations}
\crefname{figure}{Fig.}{Figs.}
\Crefname{figure}{Figure}{Figures}
\crefname{section}{Sect.}{Sects.}
\Crefname{section}{Section}{Sections}
\crefname{table}{Table}{Tables}
\crefname{appsec}{Appendix}{Appendices}

\graphicspath{{Figures/}}
\usepackage[cal=boondox]{mathalfa}

\begin{document}

\title{Efficient modeling of superconducting quantum circuits with tensor networks}
\author{Agustin Di Paolo}
\affiliation{Institut quantique \& D\'epartement de physique, Universit\'e de Sherbrooke, Sherbrooke, Qu\'ebec J1K 2R1, Canada}
\author{Thomas E. Baker}
\affiliation{Institut quantique \& D\'epartement de physique, Universit\'e de Sherbrooke, Sherbrooke, Qu\'ebec J1K 2R1, Canada}
\author{Alexandre Foley}
\affiliation{Institut quantique \& D\'epartement de physique, Universit\'e de Sherbrooke, Sherbrooke, Qu\'ebec J1K 2R1, Canada}
\author{David S\'en\'echal}
\affiliation{Institut quantique \& D\'epartement de physique, Universit\'e de Sherbrooke, Sherbrooke, Qu\'ebec J1K 2R1, Canada}
\author{Alexandre Blais}
\affiliation{Institut quantique \& D\'epartement de physique, Universit\'e de Sherbrooke, Sherbrooke, Qu\'ebec J1K 2R1, Canada}
\affiliation{Canadian Institute for Advanced Research, Toronto, ON, Canada}

\date{\today}

\begin{abstract}
We introduce an efficient tensor network toolbox to compute the low-energy excitations of large-scale superconducting  quantum  circuits  up  to a desired accuracy. We benchmark this algorithm on the fluxonium qubit, a superconducting quantum circuit based on a Josephson junction array with over a hundred junctions. As an example of the possibilities offered by this numerical tool, we compute the pure-dephasing coherence time of the fluxonium qubit due to charge noise and coherent quantum phase slips, taking into account the array degrees of freedom corresponding to a Hilbert space as large as~$15^{180}$. Our algorithm is applicable to the wide variety of circuit-QED systems and may be a useful tool for scaling up superconducting-qubit technologies.
\end{abstract}

\maketitle

\section{Introduction}
\label{s: Introduction}

Superconducting qubits are a leading platform for quantum information processing~\cite{devoret2013superconducting,arute2019quantum}. These qubits are built from superconducting quantum circuits integrating linear elements, such as capacitors and inductors, together with the only known nonlinear and nondissipative circuit component: the Josephson junction. These circuits 
operate at milliKelvin temperatures where macroscopic electromagnetic degrees of freedom associated to currents and voltages in the circuit are described quantum mechanically~\cite{devoret1995quantum,burkard2004multilevel}. In this regime, nodes (or branches) of the circuit are represented by bosonic fields with, in principle, infinite Hilbert-space dimension. The circuit topology defines linear and nonlinear interactions between these bosonic modes. Determining the low-lying excitations of the circuit in the presence of such interactions requires the diagonalization of the full circuit Hamiltonian. However, for circuits with more than a few nodes, this rapidly becomes intractable by exact diagonalization. With current devices integrating 10s~\cite{manucharyan2009fluxonium} to 100s~\cite{earnest2018realization}, 1,000s~\cite{macklin2015near} and even 10,000s~\cite{kuzmin2019quantum} Josephson junctions, finding new methods to efficiently model these devices is one of the challenges that the field is facing.

Most superconducting quantum devices operate in regimes where effective models with a reduced number of degrees of freedom are accurate enough to describe the physics of interest. However, these effective models are based on approximations that allow extracting only limited information about the system. Moreover, it is often not possible to trace back the original circuit parameters from the effective model and, when it is possible, these parameters have to be inferred indirectly from complex multivariate fits to the experimental data. This loss of information can be detrimental to circuit design.

In this work, we adapt to many-body superconducting quantum circuits a numerical tensor network method that we have introduced in Ref.~\cite{thomas2019}. We use this numerical toolbox to compute the relevant low-energy excitations of a large-scale superconducting circuit taking into consideration all of the degrees of freedom of a lumped-element model of the device. We show how this gives access to information about the system that can be used, for instance, to estimate the device coherence times from first principles.

As an example of application of this method, we consider the fluxonium qubit~\cite{manucharyan2009fluxonium}. This superconducting quantum circuit is made of a small Josephson junction shunted by an array of~$\sim100$ Josephson junctions. Because of the large number of elements in the fluxonium circuit, this qubit is an ideal testbed for our numerical approach. Moreover, solving the complete fluxonium circuit Hamiltonian is challenging due to the short- and long-range linear and nonlinear interactions of the model, which is formulated under periodic boundary conditions. To benchmark our tensor network implementation, we develop an effective model for the fluxonium qubit that captures the essential circuit details and which can easily be solved by exact diagonalization. To assert the validity of the tensor network method, we first compare results obtained with this technique to those obtained with the approximate effective model in regimes where the latter approach is expected to faithfully describe the device. We then push the tensor network method to regimes where deriving an accurate effective theory is difficult. The effective model and the tensor network toolbox are used to investigate the charge dispersion  of the fluxonium qubit in a broad range of parameters, confirming an existing theory~\cite{manucharyan2012evidence} and clarifying its regime of validity. Finally, we use the tensor network method to estimate the pure-dephasing time of a realistic fluxonium device. We provide direct numerical evidence of the potentially harmful effects of charge noise in this system for certain circuit parameters. 

This paper is organized as follows. In~\cref{s: The multi-targeted DMRG algorithm}, we summarize the tensor network method introduced in Ref.~\cite{thomas2019}. In~\cref{s: DMRG implementation of the fluxonium-qubit Hamiltonian}, we provide a tensor network implementation of the complete fluxonium-qubit Hamiltonian, describe an effective model for this qubit and compare results obtained with both approaches. \cref{ss: Charge dispersion due to coherent quantum phase slip processes} discusses the interplay between charge noise and coherent quantum phase slips in the fluxonium qubit. The main result of this section is the direct numerical evidence of the charge dispersion in fluxonium devices, supporting a previously developed theory~\cite{manucharyan2012evidence}.~\cref{s: Conclusions and outlook} is dedicated to the conclusions and to an outlook of the results of this work.

\section{The multi-targeted DMRG algorithm}
\label{s: The multi-targeted DMRG algorithm}

A useful strategy to determine the low-energy excitations of a quantum system is based on decomposing the many-body wavefunction into a series of tensors, each representing a single site (or mode). The form of the resulting wavefunction is called matrix product state (MPS) and has been known for some time~\cite{affleck1988valence}. For a review, see for instance Refs.~\cite{schollwock2005density,schollwock2011density,orus2014practical,bridgeman2017hand,baker2019m}. The tensor decomposition applies to the full many-body wavefunction
\begin{equation}
|\psi\rangle=\sum_{\{\sigma_i\}}c_{\sigma_1\sigma_2\ldots\sigma_{N_J}}|\sigma_1\sigma_2\ldots\sigma_{N_J}\rangle,
\label{eq: ManyBody Psi}
\end{equation}
where~$\sigma_i$ indexes orbitals (or levels) that belong to a finite-dimensional basis of states for the~$i$th site. For a site representing a bosonic mode, a finite-dimensional basis for this site may be defined by truncating the site's Hilbert space. The probability amplitude~$c_{\sigma_1\sigma_2\ldots\sigma_{N_J}}$ in~\cref{eq: ManyBody Psi} is interpreted as a tensor with~$N_J$ indices,~$N_J$ being the number of sites. In order to obtain a MPS representation of~$|\psi\rangle$, a series of tensor decompositions can be performed using the singular value decomposition~(SVD). The SVD decomposes a tensor into two isometries,~$U$ and~$V$, and a diagonal matrix~$D$ such that the original tensor may be reconstructed as~$UDV^\dagger$. By performing successive SVDs on the full original tensor, one obtains a site-by-site representation of the wavefunction of the form~\cite{schollwock2011density}
\begin{equation}
\begin{split}
|\psi\rangle =\sum_{\{\sigma_i\},\{a_i\}}&A^{\sigma_1}_{a_1}A^{\sigma_2}_{a_1a_2}\ldots A^{\sigma_{{N_J}-1}}_{a_{{N_J}-2}a_{{N_J}-1}}A^{\sigma_{{N_J}}}_{a_{{N_J}-1}}\\
&\times|\sigma_1\sigma_2\ldots\sigma_{N_J}\rangle,
\end{split}
\label{eq: MPS}
\end{equation}
where~$A^{\sigma_i}_{a_{i-1}a_i}$ is the tensor of the MPS associated to the~$i$th site. Here, an extra index~$a_i$ appears corresponding to a link index that connects to an adjacent site. The dimension of this additional index is known as the bond dimension and is controlled by truncating the number of nonzero singular values that are kept in the diagonal matrix~$D$ of the SVDs. Effectively, this truncation leads to a compressed representation of the many-body state, leaving out small entries of the density matrix which are unimportant to understand the physical phenomenon of interest. Physical systems that can be modeled efficiently by a MPS with a much smaller bond dimension than the full wavefunction often involve short-range interactions and low dimensions~\cite{verstraete2006matrix}. Other cases can also be captured by a MPS at the price of using a larger bond dimension~\cite{schollwock2005density,schollwock2011density}. 

\Cref{eq: MPS} is represented in the left-normalized basis where the tensor~$A$ is determined from the~$U$ tensor of the SVD. The MPS can also be written with right-normalized tensors (creating tensors from~$V^\dagger$). The most common gauge to choose is the mixed-canonical representation~\cite{schollwock2011density}. There, left- and right-normalized tensors are separated by one site where the~$D$ matrix has been contracted on to the site. This site is known as the orthogonality center, and represents the information passed between the left and right parts of the system.

In practice, the MPS is obtained by first constructing the Hamiltonian as a tensor network, known as a matrix product operator (MPO). Once the MPO is specified, an algorithm can be designed to converge from a starting initial state to the correct ground state. A well-known tensor network method to achieve this is the density matrix renormalization group (DMRG) algorithm~\cite{white1992density,white1993density}. This approach is found to be efficient for solving systems that are well captured by the MPS and can converge to the ground state in only a few iterations of the algorithm~\cite{verstraete2004renormalization,verstraete2006matrix,vidal2007entanglement}. More importantly, the complexity of this algorithm scales linearly with the number of sites, making it possible to treat systems of sizes well beyond what is possible with exact diagonalization. 

While DMRG is most commonly used to study ground-states, the analysis of superconducting quantum circuits requires us to determine several low-energy excitations. For example, in the case of a single superconducting qubit built using some large superconducting circuit, the ground state and the two first lowest energy excitations are needed to estimate the qubit frequency~$\omega_{01}$ and anharmonicity~$\omega_{12} - \omega_{01}$, where~$\hbar\omega_i$ is the energy of the~$i$th eigenstate of the circuit and~$\omega_{ij} = \omega_j-\omega_i$. If~$n_\mathrm{q}$ such qubits are integrated on a chip, the number of excitations required to characterize the device typically scales as~$n_{\mathrm{q}}^2$.

The conventional approach to compute excitations with DMRG is to add to the system Hamiltonian an energy penalty of the form~$\sum_{i\in\mathrm{ex.}}\Lambda|\psi_i\rangle\langle \psi_i|$, with~$\Lambda>0$, where $\mathrm{ex.}$ denotes a set of previously determined excitations $\{|\psi_i\rangle\}$. This energy penalty forces the previously determined low-energy excitations above the next excited state, which becomes the ground state of the modified Hamiltonian and for which standard DMRG can be run~\cite{schollwock2011density}. However, it can be noticed that this technique can miss excited states and suffers from convergence issues. 

To remedy this problem, we have derived an extension of the DMRG algorithm that includes the excitations computed directly in the Lanczos step of the algorithm~\cite{thomas2019}. We extended the original MPS to a bundled MPS, where the orthogonality center has been given an additional index that identifies excitations in the system. By attaching this additional index to the state, we can derive an efficient tensor network update at each step of the DMRG algorithm that modifies the wavefunction of each excitation until the energy is variationally minimized to the correct eigenvalue. This procedure, that we name the `multi-targeted' DMRG algorithm, is numerically stable and does not miss excitations or introduce numerical degeneracies in all tested situations. 

Indeed, we have used this method to obtain tens or hundreds of excitations simultaneously, all in a single run of the multi-targeted DMRG algorithm. This is where our newly developed technique differs significantly from the traditional DMRG approach for computing excitations, which needs to be run sequentially, once per required excitation. Furthermore, an important benefit of our multi-targeted DMRG algorithm is that the orthogonality of the computed excited states is guaranteed up to numerical precision. In contrast, in the traditional DMRG approach, the degree to which orthogonality conditions are satisfied within a set of computed eigenstates is determined by the accuracy of the associated eigenvalues. More information on the multi-targeted DMRG algorithm can be found in Ref.~\cite{thomas2019}.  

\section{DMRG implementation of the fluxonium-qubit Hamiltonian}
\label{s: DMRG implementation of the fluxonium-qubit Hamiltonian}

\begin{figure}[t!]
\includegraphics[scale=1.14]{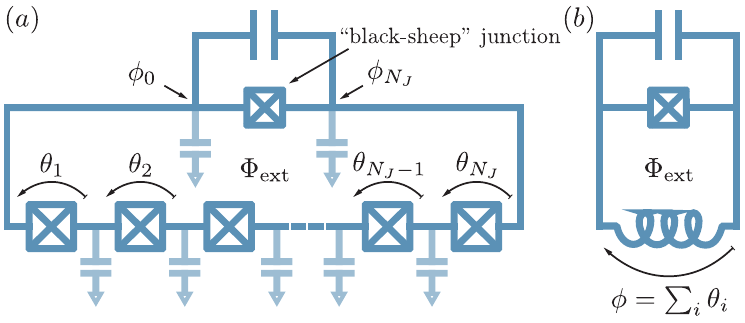}
\caption{\label{fig:circuit_design} Lumped-element model of the fluxonium qubit.~($a$) Detailed circuit scheme including a ``black-sheep'' junction (center) shunted by a capacitance (top) and a junction-array superinductance with~$N_J$ junctions (bottom). Stray capacitances to ground are depicted in a lighter shade of blue.~($b$) Effective circuit in which the junction-array is modeled as a linear inductance.~$\phi_i$ for~$i\in[0,N_J]$ denotes the superconducting phase at every circuit node, while~$\theta_i$ for~$i\in[1,N_J]$ is the phase difference at every junction of the array. The superinductance (or fluxonium) mode is defined as the phase difference across the black-sheep junction:~$\phi=\phi_0-\phi_{N_J}=\sum_{i=1}^{N_J}\theta_i$.}
\end{figure}

We choose the fluxonium qubit~\cite{manucharyan2009fluxonium} as a testbed for the multi-targeted DMRG approach. Because of its relatively complex structure, with a Hamiltonian that includes periodic boundary conditions as well as short- and long-range linear and nonlinear interactions (see~\cref{s:Fluxonium Circuit Hamiltonian}), this is an ideal test circuit for this numerical method. We note that non-multi-targeted DMRG has previously been used to study quantum phase transitions in Josephson-junction rings~\cite{chung1997superconductor,lee2003quantum} and the coherence properties of the current-mirror qubit~\cite{weiss2019spectrum}. 

The fluxonium qubit is a variation on the transmon qubit~\cite{koch2007charge} in which a large shunt inductor is added to protect the device against low frequency charge noise~\cite{koch2009a}. Recent experiments have demonstrated long coherence times with this qubit~\cite{lin2018demonstration,earnest2018realization,nguyen2018high}. The fluxonium circuit (see~\cref{fig:circuit_design}) consists of a small Josephson junction, referred to as the ``black-sheep'' junction, shunted by a superinductance, i.e.~a circuit element with effective impedance greater than the quantum of resistance~$R_Q=h/(2e)^2\simeq 6.5\,\mathrm{k}\Omega$ and self-resonance frequencies above~$10\,\mathrm{GHz}$~\cite{kitaev2006protected,brooks2013protected,bell2012quantum,masluk2012microwave,manucharyan2012superinductance}. Superinductances have been made using Josephson junction arrays~\cite{manucharyan2009fluxonium,masluk2012microwave}, high-kinetic-inductance superconductors~\cite{maleeva2018circuit,hazard2019nanowire} and granular aluminium~\cite{Grunhaupt2019,kamenov2019granular}. Superinductances are also crucial to other qubit designs such as the noise-protected~$0-\pi$ qubit~\cite{brooks2013protected,gyenis2019experimental}. While a superinductance is in principle a multimode device, it can behave as a single-mode linear inductance under appropriate conditions~\cite{masluk2012microwave,ferguson2013symmetries,viola2015collective}. The multimode structure of such a device has, however, important consequences~\cite{masluk2012microwave,hazard2019nanowire}, some of which are investigated below.

\subsection{Setting-up the multi-targeted DMRG algorithm}
\label{ss: Many-body structure of the fluxonium qubit}

With the objective of determining the low-energy excitations of the full fluxonium device shown in~\cref{fig:circuit_design}~($a$) using our multi-targeted DMRG algorithm, we first describe the associated circuit Hamiltonian. In this circuit, the black-sheep junction is described by its Josephson energy~$E_{J_{\mathrm{b}}}$ and its capacitance~$C_{J_\mathrm{b}}$ which may include a shunt capacitance. We take the superinductance to be realized by an array of Josephson junctions, with~$L_{J_i}$ and~$C_{J_i}$ being the~$i$th junction inductance and capacitance, respectively. Moreover, a ground capacitance~$C_{0_i}$ is associated to the~$i$th circuit node. In the absence of circuit element disorder, these parameters take the constant values~$L_J$,~$C_J$ and~$C_0$, respectively. We also define the junction plasma frequency~$\omega_p = 1/\sqrt{L_JC_J}$ and reduced impedance~$z=\sqrt{L_J/C_J}/R_Q$. Following the standard circuit-quantization procedure~\cite{devoret1995quantum}, the Hamiltonian of the circuit of~\cref{fig:circuit_design} takes the form (see~\cref{s:Fluxonium Circuit Hamiltonian})
\begin{equation}
H = \sum_{i=1}^{N_J} H_{0_i} + \sum_{j>i}^{N_J} \hbar g_{ij}\,n_i n_j - E_{J_{\mathrm{b}}} \cos\Bigg(\sum_{i=1}^{N_J}\theta_i + \varphi_{\mathrm{ext}}\Bigg).
\label{eq: fluxonium H}
\end{equation}
In this expression,~$H_{0_i}=4 E_{C_i}(n_i-n_{g_i})^2-E_{J_i}\cos\theta_i$ is a noninteracting (or site) Hamiltonian for the~$i$th array junction, where~$\theta_i$ is the phase difference across that junction and~$n_i$ the conjugate charge. Moreover,~$n_{g_i}$ is an offset-charge parameter,~$E_{C_i}$ is the effective charging energy of this junction and~$E_{J_i}=\varphi_0^2/L_{J_i}$ is the Josephson energy with~$\varphi_0=\Phi_0/2\pi$ where~$\Phi_0=h/2e$ is the flux quantum. In addition to the on-site energies, \cref{eq: fluxonium H} includes a bilinear interaction~$\propto n_i n_j$ arising from the ground, black-sheep and array-junction capacitances, that couples the sites with comparable strength and all-to-all connectivity (see~\cref{s:Fluxonium Circuit Hamiltonian}). Furthermore, the last term of~\cref{eq: fluxonium H} is a nonlocal interaction that depends on the external flux~$\Phi_\mathrm{ext}=\varphi_0\varphi_\mathrm{ext}$ and which results from the strongly nonlinear Josephson potential of the black-sheep junction. Because~\cref{eq: fluxonium H} includes a very large number of degrees of freedom and is therefore difficult to work with, this Hamiltonian is typically not directly employed in the literature to describe the fluxonium qubit. Instead, fluxonium devices are usually modeled by a phenomenological Hamiltonian that incorporates a single bosonic degree of freedom,~$\phi = \textstyle\sum_{i=1}^{N_J}\theta_i$, known as superinductance or fluxonium mode~\cite{manucharyan2009fluxonium}. 

To obtain the low-energy excitations of Eq.~\eqref{eq: fluxonium H} by means of a tensor network method, and in this way go beyond the usual effective model, the circuit Hamiltonian must first be converted to its matrix product operator form. Crucially, we noticed that the long-range cosine interaction is ideally suited to matrix product states and operators, preventing an increase of the bond dimension with the number of sites. This observation is one of the key findings of our work and extends to all circuit-QED Hamiltonians, from lumped-element models to black-box-quantization~\cite{nigg2012black,bourassa2012josephson} and energy-participation-ratio~\cite{minev2019catching} formalisms. Indeed, we have successfully implemented a wide variety of such models and circuit Hamiltonians, results that will be  reported elsewhere. On the other hand, the all-to-all capacitive interaction in~\cref{eq: fluxonium H} does not have an efficient MPO representation. However, this unfavorable interaction does not prevent an efficient implementation of the multi-targeted DMRG algorithm, as the results that are presented below are obtained with a relatively small bond dimension using MPO compression techniques~\cite{hubig2017generic}. The efficient matrix-product-operator representation of the black-sheep Josephson potential in~\cref{eq: fluxonium H}, and the possibility of handling an arbitrary capacitive coupling Hamiltonian by compression techniques, makes our DMRG implementation readily applicable to the wide variety of circuit-QED setups. 

\subsection{Effective single-mode theory}

To assert the validity of our DMRG method, we derive in~\cref{s:Effective models for the fluxonium qubit} an effective single-mode theory from~\cref{eq: fluxonium H} that can be solved by exact diagonalization, and which goes beyond the standard treatment found in the literature. Under approximations controlled by the parameter regime of the device, we arrive at the Hamiltonian
\begin{equation}
H' = 4E_C n^{\prime 2} - N_J^2E_{L}\cos(\phi'/N_J) - E_{J} \cos(\phi' + \varphi_{\mathrm{ext}}),
\label{eq: effective fluxonium H}
\end{equation}
where the mode described by~$\phi'$ is closely related to the superinductance (or fluxonium) mode~$\phi$, and~$n'$ the conjugate charge. Here,~$E_C$,~$E_L$ and~$E_J$ are, respectively, effective capacitive, inductive and Josephson energies obtained from the classical normal-mode structure of the circuit. If the ground capacitances~$C_{0_i}$ for~$i\in[1,N_J]$ can be neglected, then~$\phi' = \phi$ and~$n' = n = N_J^{-1}\textstyle\sum_{i=1}^{N_J} n_i$, where~$n$ is the conjugate charge operator to~$\phi$. Otherwise, the~$\phi'$ mode includes corrections to~$\phi$ that are linear in~$C_0$. 

Although in the limit of large~$N_J$~\cref{eq: effective fluxonium H} reduces to the usual effective model for the fluxonium-qubit~[see~\cref{fig:circuit_design}~($b$)]~\cite{manucharyan2009fluxonium}, the parameters of~\cref{eq: effective fluxonium H} capture the full circuit's capacitance network and contain important corrections due to the nonlinearity of the array junctions. These corrections can lead to significant frequency shifts of the qubit transitions (see~\cref{ss:Exploration of various parameter regimes}). Crucially, because of its single-mode nature,~\cref{eq: effective fluxonium H} can easily be diagonalized numerically by truncating the Hilbert space of the~$\phi'$ mode to finite dimension. 

\subsection{Comparison}

Having derived the effective model of~\cref{eq: effective fluxonium H} which will be used as a benchmark, we are now in a position to demonstrate the results of our DMRG approach and to explore the capabilities of this method. To this end, we consider a device in the `heavy fluxonium' regime~\cite{earnest2018realization,lin2018demonstration,hazard2019nanowire} with a large shunt capacitance and a superinductance made of~$N_J=120$ identical junctions where~$\omega_p/2\pi=25\,\mathrm{GHz}$ and~$z=0.03$~\cite{masluk2012microwave}. See~\cref{ss:Qualitative regimes of the fluxonium qubit} for a qualitative description of the different regimes of the fluxonium qubit Hamiltonian. Each junction is modeled as a multilevel system using the~$15$ lowest energy eigenstates of the site Hamiltonian~$H_{0_i}$. We find that for low-impedance junctions, the site eigenbasis requires a smaller number of states to avoid truncation errors as compared to other local bases such as the charge basis. The DMRG implementation is thus defined in a product basis of local wavefunctions spanning a many-body Hilbert space as large as $15^{120}$ and that has, a priori, no built-in information about collective modes of the system. Importantly, this choice of basis also makes our treatment readily extensible to other superconducting quantum circuits. 

\begin{figure}[t!]
\includegraphics[scale=1.]{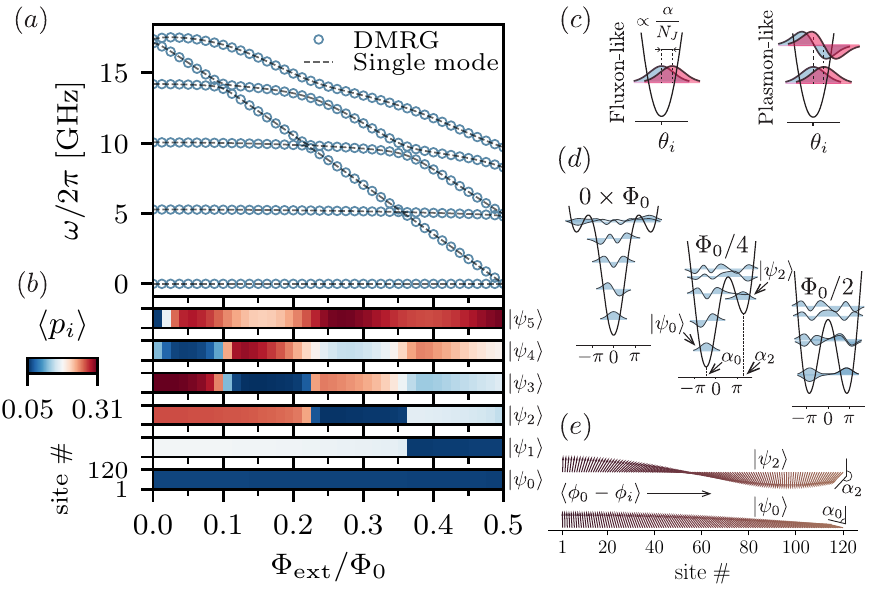}
\caption{\label{fig:fluxo_spectrum} A~$120$-junction superinductance heavy fluxonium as a function of~$\Phi_{\mathrm{ext}}$.~($a$) Energy spectrum of the Hamiltonians in~\cref{eq: fluxonium H} (DMRG) and~\cref{eq: effective fluxonium H} (single mode).~($b$) Mean photon-number population of the array Josephson junctions (sites) for every eigenstate~$|\psi_k\rangle$ of the fluxonium circuit.~($c$) Single-junction picture of fluxon- and plasmon-like excitations.~($d$) Schematic of the effective potential energy and wavefuctions of the single-mode Hamiltonian for~$\Phi_{\mathrm{ext}}\in\{0,\Phi_0/4,\Phi_0/2\}$.~($e$) Expectation value of the phase operator at every circuit node of the superinductance for the fluxonium eigenstates labeled by~$|\psi_0\rangle$ and~$|\psi_2\rangle$. Circuit parameters:~$C_{J_{\mathrm{b}}}=40\,\mathrm{fF}$,~$E_{J_{\mathrm{b}}}/h=7.5\,\mathrm{GHz}$,~$C_J\simeq 32.9\,\mathrm{fF}$ and~$L_J\simeq 1.23\,\mathrm{nH}$ (from~$\omega_p/2\pi=25\,\mathrm{GHz}$ and~$z = 0.03$~\cite{masluk2012microwave}) and~$C_0=0$. Single-mode model parameters:~$E_C/h\simeq 0.48\,\mathrm{GHz}$,~$E_L/h\simeq 1.27\,\mathrm{GHz}$ (i.e.~$L\simeq 129.1\,\mathrm{nH}$) and~$E_J=E_{J_{\mathrm{b}}}$.}
\end{figure}

\Cref{fig:fluxo_spectrum}~($a$) shows the energy spectrum of the fluxonium device of~\cref{fig:circuit_design} for both multi-targeted DMRG [\cref{eq: fluxonium H}, light-blue circles] and exact diagonalization of the effective single-mode theory [\cref{eq: effective fluxonium H}, black dashed lines] as a function of the external flux~$\Phi_{\mathrm{ext}}$. We find excellent agreement between these two independent models. Importantly, this 
observation extends to all systems sizes and parameter sets that we have 
tested, from a few-sites fluxonium-like device to circuits with more than~$200$ junctions. These results provide supporting evidence of a successful DMRG implementation of the fluxonium qubit Hamiltonian. Moreover, this motivates applying the DMRG technique in regimes of parameters where deriving an effective model is not possible. Further numerical evidence is presented in~\cref{s:Effective models for the fluxonium qubit}.

\subsection{Exploring the DMRG results}

In addition to computing global properties of the circuit, such as its energy spectrum, the multi-targeted DMRG algorithm also gives access to local site properties and $n$-body correlators. These operators can give insights into the many-body structure of the fluxonium eigenstates. The purpose of this section is to motivate the use of our DMRG algorithm to explore some of these quantities.

As an example application,~\cref{fig:fluxo_spectrum}~($b$) shows the mean photon-number population~$\langle p_i\rangle = \langle \psi_k| H_{0_i}|\psi_k\rangle/\hbar\omega_p$ of the~$i$th site, for all sites ($i\in[1,120]$, vertical axis of each of the 6 density plots) as a function of~$\Phi_{\mathrm{ext}}$. These expectation values are computed for a given eigenstate~$|\psi_k\rangle$ of the full fluxonium circuit, from the ground state ($k=0$, bottom density plot) to the~$5$th excited state ($k=5$, top density plot). Because of the absence of circuit-element disorder in these simulations, the results do not show any variations with site number. We observe that the photon-number population of the array junctions is relatively low for the ground state. The same is true for some excited states whose energies change rapidly with the external flux (fluxons). Note that energies are given with respect to the ground state energy, which is chosen to be always $0$. In other words, the energy of the $i$th excited state as illustrated in \cref{fig:fluxo_spectrum} ($a$) corresponds to that of the transition $|\psi_0\rangle\to |\psi_i\rangle$. Moreover, we note that the photon-number population of the array junctions is relatively high for excited states that have a weak frequency dispersion as a function of $\Phi_{\mathrm{ext}}$ (plasmons). We interpret these results with the help of~\cref{fig:fluxo_spectrum}~($c$), which illustrates a portion of the local Josephson potential of an array junction and its single-site wavefunctions. From the point of view of this site (left panel), a fluxon state~$|\psi_k\rangle$ involves a small displacement by~$\alpha_k/N_J$ of the site's wavefunction (red) away from its noninteracting ground state position (light blue), where $\alpha_k$ is a real number. With the current operator associated to the~$i$th junction given by~$I_i = I_{c}\sin\theta_i$ where~$I_{c}$ is critical current, this displacement results in a circulating current for $\alpha_k\neq 0$. In addition to this mean-field displacement, plasmon states involve non-negligible population of the sites' excited states, as shown in~\cref{fig:fluxo_spectrum}~($c$)~[right panel].

The above interpretation becomes clearer by considering the effective potential and wavefunctions obtained from the single-mode effective Hamiltonian~\cref{eq: effective fluxonium H}, as shown in~\cref{fig:fluxo_spectrum}~($d$) for~$\Phi_{\mathrm{ext}}\in\{0,\Phi_0/4,\Phi_0/2\}$. The shape of the effective potential is determined by the cosine potential of the black-sheep junction and the inductive energy~$-N_J^2E_{L}\cos(\phi'/N_J)\simeq E_L \phi'^2/2$ of the array. While fluxon states correspond in this picture to the lowest energy eigenstates associated to the local minima of the effective potential, plasmon states correspond to intra-well excitations (see also~\cref{ss:Qualitative regimes of the fluxonium qubit}). The effective model potential connects to that of~\cref{fig:fluxo_spectrum}~($c$) by noticing that~$\langle\psi_k|\phi|\psi_k\rangle\equiv\langle\psi_k|\textstyle\sum_{i=1}^{N_J}\theta_i|\psi_k\rangle=\alpha_k$ for an excitation $|\psi_k\rangle$ localized in a single potential well. Thus, in this case, the displacement of the sites' wavefunctions adds to a collective value $\alpha_k$ that approximately coincides with the position of a local minimum of the effective potential. This is examined further in \cref{fig:fluxo_spectrum}~($e$), which shows the expectation value of the phase drop $\phi_0-\phi_i \equiv \sum_{j=1}^{i}\theta_j$, obtained from DMRG and plotted as a function of the site number for the fluxon states~$|\psi_0\rangle$ and~$|\psi_2\rangle$ at~$\Phi_{\mathrm{ext}}=\Phi_0/4$ in \cref{fig:fluxo_spectrum}~($d$)~[middle panel]. In this figure, the expectation value~$\langle\psi_k|(\phi_0-\phi_i)|\psi_k\rangle$ is represented by the angle between the direction of a vector localized on the $i$th site with respect to the vertical direction. Thus, the total angle between the vectors belonging to the first and last sites can be identified with the positions of the local minima~$\alpha_0$ and~$\alpha_2$ of the effective potential of~\cref{eq: effective fluxonium H}. 

Overall,~\cref{fig:fluxo_spectrum} shows that the multi-targeted DMRG algorithm  correctly reproduces the results of the effective single-mode theory. It can also provide information that is not accessible from this theory. This comparison provides solid evidence of a correct DMRG implementation of the full circuit Hamiltonian of the fluxonium qubit. It also suggests that other circuit Hamiltonians can benefit from this numerical method. Moreover, the local physical quantities such as those illustrated in~\cref{fig:fluxo_spectrum}~($b$), contain information about the energy-participation ratio of all circuit components for a given collective excitation. This information could be used to identify limiting dissipation channels and to understand the effect of circuit-element disorder. We return to these aspects in~\cref{s: Conclusions and outlook}.

\section{Charge dispersion and coherence time}
\label{ss: Charge dispersion due to coherent quantum phase slip processes}

We now proceed with a concrete application that shows how 
our DMRG implementation can be leveraged to produce coherence-time estimates from first principles. In particular, we are interested in quantifying the coherence time of the fluxonium due to the combined effect of charge noise and coherent quantum phase slips~\cite{manucharyan2012superinductance,manucharyan2012evidence}. 

\subsection{Charge dispersion}
\label{sss: Charge dispersion}

In the fluxonium qubit, the black-sheep junction acts as a weak link that couples flux states of the superconducting loop. This mechanism makes quantum control of the flux degree of freedom possible but can also be a source of errors. In a semiclassical picture, the rate at which a quantum of flux can tunnel in and out of the loop through the black-sheep junction is proportional to the junction impedance, while the energy cost associated to the addition of a quantum of flux to the loop scales as~$1/L$. Since the tunneling of a flux quantum corresponds to a change of~$2\pi$ in the phase of the superconducting order parameter, this phenomenon is known as coherent quantum phase slip (CQPS)~\cite{matveev2002persistent,mooij2005phase,mooij2006superconducting,hriscu2011coulomb,manucharyan2012evidence,rastelli2013quantum,susstrunk2013aharonov}. In experiments, fluxonium devices exploit a wide range of black-sheep junction impedances, ranging from relatively small in the heavy-fluxonium~\cite{earnest2018realization,lin2018demonstration,hazard2019nanowire}, to moderate in the fluxonium~\cite{manucharyan2009fluxonium,manucharyan2012evidence} and to large values for the light-fluxonium~\cite{pechenezhskiy2019quantum}. See~\cref{ss:Qualitative regimes of the fluxonium qubit} for a qualitative discussion of these parameter regimes. Ideally, the total amplitude for CQPS events is largely dominated by the contribution from the black-sheep junction. However, if the impedance of the array junctions is large enough, the added CQPS amplitude due to the superinductance can be non-negligible. In this limit, the junction array may be regarded as a ``slippery'' superinductance~\cite{manucharyan2012superinductance}.

Reference~\cite{manucharyan2012evidence} introduced an effective model describing the effect of CQPS events occurring in the superinductance of a fluxonium qubit. In this model, CQPS events due to the black-sheep junction are captured by a phenomenological single-mode fluxonium qubit Hamiltonian similar in spirit to~\cref{eq: effective fluxonium H}. On the other hand, CQPS due to the superinductance enter in the effective Hamiltonian via the external flux. More precisely, the parameter~$\Phi_\mathrm{ext}$ in~\cref{eq: effective fluxonium H} is replaced by~$\Phi_\mathrm{ext} + m\,\Phi_0$, where~$m$ is an integer-valued number operator that counts the number of CQPS in the superinductance. Since a CQPS event at any junction of the superinductance leads to a jump~$m\to m\pm 1$, it can be interpreted as a~$2\pi$ phase bias on~$\phi$.

To quantify the total CQPS amplitude resulting from the superinductance, we consider a realistic model of this composite circuit element with its~$N_J$ islands and their independent offset charges [see~\cref{fig:circuit_design}~($a$)]. As a consequence of the Aharonov-Casher effect, the flux-tunneling amplitude at a given array junction has a well-defined phase given by the offset-charge~$n_{g_i}$ associated to that junction~\cite{matveev2002persistent,friedman2002aharonov,pop2012experimental,manucharyan2012evidence,susstrunk2013aharonov,bell2016spectroscopic}. By adding coherently the contributions from the~$N_J$ array junctions, the total CQPS amplitude (excluding the black-sheep junction) takes the form~$E_S = \textstyle\sum_{i=1}^{N_J}\epsilon_{0_i} e^{i2\pi n_{g_i}}$, where 
\begin{equation}
\epsilon_{0_i}=8\sqrt{2}\,\hbar\omega_{p_i}\exp(-4/\pi z_i)/\sqrt{\pi z_i},
\label{eq: charge dispersion}
\end{equation} 
determines the charge dispersion of the ground state energy of the transmon Hamiltonian~$H_{0_i}$ in terms of the reduced impedance~$z_i$ and plasma frequency~$\omega_{p_i}$ of the~$i$th array junction~\cite{matveev2002persistent,koch2007charge,catelani2011relaxation,manucharyan2012superinductance}. Importantly, this result only holds in the low-impedance limit ($z_i\ll 1$).

CQPS events in the superinductance can then be described by a phenomenological flux-tunneling Hamiltonian of the form~$H_{\mathrm{CQPS}}=(E_S\,m^{-} + E_S^*\,m^{+})/2$, where the operator~$m^{-}$ [$m^{+}=(m^{-})^{\dagger}$] removes (adds) a single flux quantum from the loop through any of the array junctions. In the limit of rare CQPS,~$|E_S|\ll E_L$,~$H_{\mathrm{CQPS}}$ can be regarded as a small perturbation to the fluxonium Hamiltonian. In this situation, first-order perturbation theory predicts a shift~$\delta\omega_{ij} = \mathrm{Re}[E_S](\langle T \rangle_j-\langle T \rangle_i)/\hbar$ of the qubit's~$i\to j$ transition frequency, where~$T=\exp(-i 2\pi n)$ is a~$2\pi$-displacement operator whose expectation values are computed using the unperturbed eigenstates~$\{|\psi_{i}\rangle\}$ with~$m=0$~\cite{manucharyan2012evidence}. For a homogeneous array ($\epsilon_{0_i}\equiv\epsilon_0$ for~$i\in[1,N_J]$), one has~$-N_J\epsilon_{0}\leq\mathrm{Re}[E_S]\leq N_J \epsilon_{0}$, and the total charge dispersion of the qubit transition frequency is
\begin{equation}
|\Delta\omega_{01}|=2N_J\epsilon_0|\langle T \rangle_1-\langle T \rangle_0|/\hbar.
\label{eq: charge dispersion total}
\end{equation} 
As the classical flux states of the loop are degenerate at~$\Phi_{\mathrm{ext}}=\Phi_0/2$, the effect of a nonzero~$E_S$ is stronger close to this flux bias. 

\begin{figure}[t!]
\includegraphics[scale=1.]{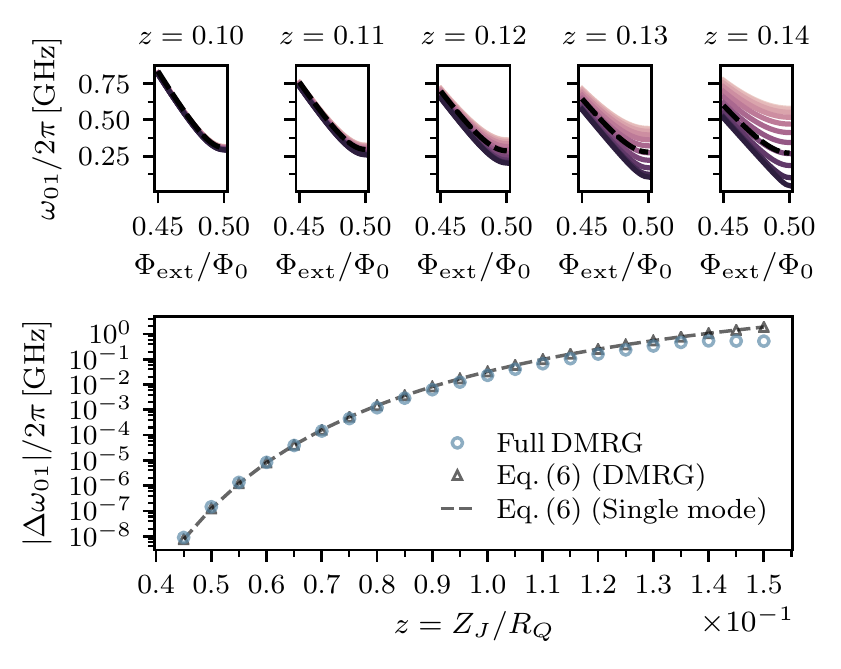}
\caption{\label{fig:vlad} Charge dispersion of a~$40$-junction superinductance fluxonium qubit as a function of the reduced impedance of the array junctions. Top panel: Broadening of the fluxon transition around~$\Phi_{\mathrm{ext}}=\Phi_0/2$ for~$n_g\in[0,0.5]$. Color lines are obtained using the multi-targeted DMRG algorithm while dashed black lines correspond to estimations using the single-mode Hamiltonian~\cref{eq: effective fluxonium H}. Bottom panel: Total charge dispersion of the fluxon transition at~$\Phi_{\mathrm{ext}}=\Phi_0/2$ according to the DMRG calculation (circles) contrasted to the prediction of~\cref{eq: charge dispersion total} with matrix elements evaluated by means of DMRG (triangles) or the single-mode model (dashed lines). Parameters:~$C_{J_{\mathrm{b}}}=7.5\,\mathrm{fF}$,~$E_{J_{\mathrm{b}}}/h=8.9\,\mathrm{GHz}$,~$\omega_p/2\pi=12.5$ and~$C_0=0$, according to Ref.~\cite{manucharyan2012evidence}.}
\end{figure}

\Cref{fig:vlad} shows the charge dispersion of the fluxon transition of a fluxonium device with parameter values chosen to be as close as possible to those of the experiment of Ref.~\cite{manucharyan2012evidence}. The top panel shows the qubit transition frequency as a function of the external flux close to~$\Phi_{\mathrm{ext}}=\Phi_0/2$ for different values of the offset charge~$n_{g_i}\equiv n_g$, assumed to be the same on every junction of the array. Each sub-panel shows the DMRG results for a given value of the array-junction impedance. The lightest (darkest) transition in purple corresponds to~$n_{g}=0$ ($n_{g}=0.5$). Since~$n_g=0.5$ is a charge degeneracy point of the single-array-junction Hamiltonian $H_{0_i}$, Cooper-pair transport between the circuit islands is relatively easier, leading to a stronger flux dispersion in comparison to the case of~$n_g=0$ \cite{fazio2001quantum}. Dashed black lines show the qubit transition according to~\cref{eq: effective fluxonium H} which does not have an offset-charge parameter. Note that the offset-charge dependence of the CQPS tunneling energy leads to constructive ($|E_S|>0$) and destructive ($E_S\to 0$) interference of CQPS events. 

Qualitatively, charge dispersion increases rapidly with~$z$ due to the exponential scaling of~\cref{eq: charge dispersion}. This is best illustrated by the bottom panel of~\cref{fig:vlad}, which shows the charge dispersion for~$\Phi_{\mathrm{ext}}=\Phi_0/2$ as a function of~$z$. Light-blue circles (Full DMRG) correspond to a fully numerical estimation using DMRG for which the charge dispersion is computed by taking the difference between the energy of the fluxon transition for~$n_g=0$ and~$n_g=0.5$. Black triangle symbols [\cref{eq: charge dispersion total} (DMRG)] are the result of~\cref{eq: charge dispersion total} for which the matrix elements are evaluated using the eigenstates obtained from DMRG for~$n_g=0$. The black dashed line [\cref{eq: charge dispersion total} (Single mode)], in contrast, is obtained by evaluating the matrix elements using the single-mode Hamiltonian~\cref{eq: effective fluxonium H}. We find no significant difference between the DMRG [\cref{eq: charge dispersion total} (DMRG)] and the single-mode [\cref{eq: charge dispersion total} (Single mode)] implementations of~\cref{eq: charge dispersion total}, with both approaches showing a small but clearly visible deviations from the results obtain from fully numerical DMRG estimation (Full DMRG) at large~$z$. 

Indeed, we observe a remarkable agreement between the estimation of the total charge dispersion from fully numerical DMRG and that predicted by~\cref{eq: charge dispersion total}, up to array-junction impedances as high as~$z\simeq~0.1$. This provides evidence in support of the theoretical model introduced in Ref.~\cite{manucharyan2012superinductance}. Although not visible in~\cref{fig:vlad}, small deviations between the fully numerical DMRG estimation and those based on~\cref{eq: charge dispersion total} are present for~$z\lesssim 0.06$. The largest truncation error for all simulations in~\cref{fig:vlad} is of order~$10^{-11}$, and the error tolerance on the eigenvalues are set to~$10^{-12}$, guaranteeing the convergence of the fully numerical DMRG results to the same accuracy. DMRG being a variational method, we have verified that the convergence to the reported accuracy is also well behaved. We noticed deviations of the same order of magnitude between the fully numerical DMRG estimation and the prediction of~\cref{eq: charge dispersion total} for devices with different sets of circuit parameters. 

On the other hand, the large relative difference between the full numerical multi-targeted DMRG estimation and those based on~\cref{eq: charge dispersion total} in the range of~$z\gtrsim 0.1$ is expected. Indeed, in this regime,~\cref{eq: charge dispersion} and the assumption that~$|E_S|\ll E_L$ are both no longer valid~\cite{manucharyan2012superinductance}. Therefore,~$z\gtrsim 0.1$ is a regime in which the DMRG method is at a clear advantage over effective theories.

\subsection{Coherence-time estimations}
\label{sss: Coherence-time estimations}

Because of unavoidable charge noise, the value of~$\delta\omega_{ij}$ fluctuates in time, leading to broadening of the qubit transition. For large charge dispersion, this effect can severely compromise qubit coherence. This observation is the basis of the experimental study of Ref.~\cite{manucharyan2012evidence}, where the reduction of the qubit coherence time around the flux sweet spot is taken as indirect evidence of CQPS events in the ``slippery'' superinductance. In support of the experimental observation and as a further example of the power of the multi-targeted DMRG algorithm, we show below that full DMRG simulation of a device with similar circuit parameters to those reported in Ref.~\cite{manucharyan2012evidence} predicts the pure-dephasing coherence times to be dominated by the combined effect of charge noise and CQPS around~$\Phi_{\mathrm{ext}}=\Phi_0/2$. Moreover, the coherence-time values that we obtain with this method result very close to those measured experimentally.

In order to estimate the coherence times, we follow closely Ref.~\cite{manucharyan2012evidence} assuming that the variables~$n_{g_i}$ are independent and randomly distributed. The probability density function of~$\mathrm{Re}[E_S]$ can then be approximated by a Gaussian distribution with zero mean and standard deviation~$\sqrt{N_J/2}\,\epsilon_0$~\cite{manucharyan2012evidence}. Following this expression, the effective broadening of the qubit transition scales as~$\sqrt{N_J}$, something which translates to the pure-dephasing rate $1/T_{\varphi,\mathrm{CQPS}}=|\Delta\omega_{01}|/4\sqrt{N_J}$~\cite{manucharyan2012evidence,manucharyan2012superinductance}. To identify the domimant dephasing mechanism, we compare this timescale to that expected for~$1/f$ flux noise by deriving in~\cref{s:Multilevel pure-dephasing master equation for flux noise} a multilevel pure-dephasing master equation of the form 
\begin{equation}
\begin{split}
\partial_t\rho &= \sum_{k} \Gamma_{\varphi}^{kk}\,\mathcal{D}[\sigma_{kk},\sigma_{kk}]\,\rho \\
& + \sum_{k>l} \Gamma_{\varphi}^{kl}\,\Big(\mathcal{D}[\sigma_{kk},\sigma_{ll}]+\mathcal{D}[\sigma_{ll},\sigma_{kk}]\Big)\rho,
\end{split}
\label{eq: pure dephasing lindbladian main}
\end{equation}
where~$\Gamma_{\varphi}^{kl}$ are time-dependent pure-dephasing rates proportional to the~$1/f$ flux noise amplitude,~$\sigma_{kl}=|\psi_k\rangle\langle\psi_l|$, and~$\mathcal{D}[x,y]\,\rho = x\rho y^{\dagger}-\{y^{\dagger}x,\rho\}/2$ is a generalized dissipator operator. By integrating~\cref{eq: pure dephasing lindbladian main}, we define the flux-noise coherence time~$T_{\varphi,\mathrm{Flux}}$ by the implicit equation~$\rho_{01}(T_{\varphi,\mathrm{Flux}})/\rho_{01}(0)=1/e$ that we solve numerically.

\begin{figure}[t!]
\includegraphics[scale=1.]{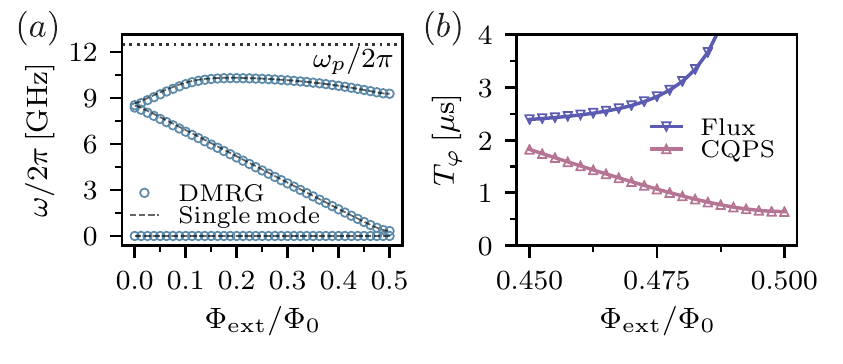}
\caption{\label{fig:vlad_device} Coherence time of a~$40$-junction superinductance fluxonium qubit.~($a$) Energy spectrum according to DMRG and single-mode estimations as a function of~$\Phi_{\mathrm{ext}}$. The black dotted line corresponds to the plasma frequency of the array junctions.~($b$) Pure-dephasing coherence times for flux and charge (CQPS) noise as obtained from DMRG. Parameters:~$C_{J_{\mathrm{b}}}=7.5\,\mathrm{fF}$,~$E_{J_{\mathrm{b}}}/h=8.9\,\mathrm{GHz}$,~$z=0.09$,~$\omega_p/2\pi=12.5$ and~$C_0=0$, extracted from Ref.~\cite{manucharyan2012evidence}. The~$1/f$ flux-noise amplitude is taken to be~$A_{\Phi}=1.2\,\mu\Phi_0$, which is a conservative value~\cite{koch2007charge}.}
\end{figure}

\Cref{fig:vlad_device}~($a$) shows the energy spectrum of the simulated device as a function of the external flux, results that should be compared to those of Ref.~\cite{manucharyan2012evidence}. In contrast to the results in~\cref{fig:fluxo_spectrum}~($a$), the difference between the DMRG and single-mode simulations for the parameters of Ref.~\cite{manucharyan2012evidence} is sightly more noticeable due to the low plasma frequency of the array junctions $\omega_p/2\pi=12.5\,\mathrm{GHz}$, around which~$\sim 40$ other additional circuit modes lie~\cite{hazard2019nanowire}. This makes any single-mode approximation invalid, except at low frequencies. Furthermore,~\cref{fig:vlad_device}~($b$) shows the estimation of the device's coherence times using only the results from DMRG as a function of the external flux and close to the bias point~$\Phi_{\mathrm{ext}}=\Phi_0/2$. We find values which are very similar to the experimental observation (see Fig.~4 in Ref.~\cite{manucharyan2012evidence}), thus providing further numerical evidence of the combined effects of charge noise and CQPS. This mechanism dominates over flux noise close to the device's flux sweet spot, resulting in sub-$\mu$s coherence times for the device parameters of Ref.~\cite{manucharyan2012evidence}, in agreement with the experimental observations.

Combined, the results of~\cref{fig:vlad} and~\cref{fig:vlad_device} illustrate the rich interplay between charge noise and CQPS in the fluxonium architecture. Added to the improved simulation capabilities provided by the multi-targeted DMRG algorithm, these findings motivate a systematic experimental study to understand these effects further.

\section{Conclusions and outlook}
\label{s: Conclusions and outlook}

We have developed a multi-targeted DMRG algorithm to simulate large-scale superconducting quantum devices. As an example, we have applied this numerical technique to the fluxonium qubit. The fluxonium circuit integrates a large number of degrees of freedom with linear and nonlinear short- and long-range interactions that are subject to periodic boundary conditions. Combined, these features make this model a challenging target for our DMRG algorithm. To assert the validity of the DMRG simulations, we have developed a detailed single-mode theory for the fluxonium qubit. Finally, we have employed DMRG to investigate the combined effect of charge noise and coherent quantum phase slips in the fluxonium qubit, confirming the theoretical model introduced in Ref.~\cite{manucharyan2012evidence} and reproducing some of the experimental findings of that work. 

Having access to the expectation values of local and of~$n$-body operators makes it possible to investigate the many-body properties of superconducting quantum circuits. This could help, for instance, in finding new approaches to encode quantum information nonlocally in protected subspaces by exploiting entanglement in these systems. Moreover, local information of large-scale superconducting quantum circuits may be used to evaluate the impact of dissipation channels and circuit-element disorder. This might also lead to a more detailed understanding of dissipation and decoherence mechanisms. Our numerical approach also has the potential to enable advancements in several areas of superconducting-qubit research. In particular, we envision future applications to the analysis of multi-qubit devices and the design of scalable superconducting-qubit architectures.

\section*{Acknowledgments}

We thank J. Cohen, A. Gyenis, C. Leroux, Z. Minev and A. Petrescu for stimulating discussions. T.E.B.~thanks the support of the Postdoctoral Fellowship from Institut quantique and support from Institut Transdisciplinaire d'Information Quantique (INTRIQ). This work was undertaken in part thanks to funding from NSERC, the Canada First Research Excellence Fund and the ARO grant No. W911NF-17-S-0008.

\appendix
\section{Fluxonium Circuit Hamiltonian}
\label{s:Fluxonium Circuit Hamiltonian}

\subsection{Hamiltonian without gate voltages}
\label{ss:Zero charge dispersion}

We derive the circuit Hamiltonian used in the DMRG calculations presented in the main text. We consider a fluxonium device where a black-sheep Josephson junction with capacitance~$C_{J_{\mathrm{b}}}$ (including both shunt and junction capacitances) and Josephson energy~$E_{J_{\mathrm{b}}}$ is shunted by a superinductance made of~$N_J$ junctions, each of capacitance~$C_{J_i}$ and energy~$E_{J_i}$ with~$i \in [1,N_J]$. We moreover assume that each circuit node of the superinductance is connected to ground by a stray capacitance~$C_{0_i}$. The~$N_J+1$ node flux (phase) variables of the circuit are denoted by~$\Phi_i$ ($\phi_i=\Phi_i/\varphi_0$), where~$\varphi_0=\hbar/2e$ is the reduced quantum of magnetic flux and~$i\in[0,N_J]$ [see also \cref{fig:circuit_design}~($a$)]. The circuit Lagrangian can then be written as~\cite{devoret1995quantum}
\begin{equation}
\begin{split}
L(\boldsymbol{\Phi},\boldsymbol{\dot{\Phi}}) &= \frac{C_{J_{\mathrm{b}}}}{2}(\dot{\Phi}_{N_J}-\dot{\Phi}_0)^2 + \sum_{i=1}^{N_J}\frac{{C_J}_{i}}{2}(\dot{\Phi}_{i}-\dot{\Phi}_{i-1})^2\\
& + \sum_{i=0}^{N_J} \frac{C_{0_i}}{2}\dot{\Phi}^2_i + \sum_{i=1}^{N_J} E_{J_i}\cos\Big[({\Phi_i}-{\Phi_{i-1}})/\varphi_0\Big]\\
& + E_{J_{\mathrm{b}}} \cos\Big[(\Phi_{N_J}-\Phi_0 + {\Phi_{\mathrm{ext}}})/{\varphi_0}\Big],
\end{split}
\label{eq: superfluxo lagrangian}
\end{equation}
where~$\Phi_{\mathrm{ext}}$ is the flux through the circuit loop. A more convenient basis is defined by the flux variables~$\Theta_i=\Phi_{i-1}-\Phi_i$ for~$i\in[1,N_J]$ and the cyclic mode~$\Sigma = \sum_{i=0}^{N_J}\Phi_i$. The relation between the new modes and the original node fluxes can be expressed concisely by~$\boldsymbol{\Theta}=\boldsymbol{\mathrm{R}}\cdot\boldsymbol{\Phi}$, where~$\boldsymbol{\Theta} = (\Theta_1,\dots,\Theta_{N_J},\Sigma)^T$,~$\boldsymbol{\Phi} = (\Phi_0,\dots,\Phi_{N_J})^T$ and~$\boldsymbol{\mathrm{R}}$ is the~$N_J+1\times N_J+1$ matrix
\begin{equation}
\boldsymbol{\mathrm{R}} = 
\begin{pmatrix}
1 & -1 & 0 & \cdots & \cdots & \cdots & 0\\
0 & 1 & -1 & 0 & \cdots & \cdots & 0\\
0 & 0 & 1 & -1 & 0 & \cdots & 0\\
\vdots & \vdots & \ddots & \ddots & \ddots & \ddots & \vdots\\
0 & \cdots & \cdots & 0 & 1 & -1 & 0\\
0 & \cdots & \cdots & \cdots & 0 & 1 & -1\\
1 & 1 & 1 & \cdots & \cdots & 1 & 1\\
\end{pmatrix}.
\label{eq: R matrix}
\end{equation}
Under this change of basis,~\cref{eq: superfluxo lagrangian} becomes
\begin{equation}
\begin{split}
L(\boldsymbol{\Theta},\boldsymbol{\dot{\Theta}}) &= \dot{\Theta}^T\cdot \frac{\boldsymbol{C}_{\Theta}}{2} \cdot \dot{\Theta} + \sum_{i=1}^{N_J} E_{J_i}\cos(\Theta_i/\varphi_0) \\
&+ E_{J_{\mathrm{b}}} \cos\Bigg[\Big(\sum_{i=1}^{N_J}\Theta_i + {\Phi_{\mathrm{ext}}}\Big)/{\varphi_0}\Bigg],
\end{split}
\label{eq: superfluxo lagrangian on difference-mode basis}
\end{equation}
where~$\boldsymbol{C}_{\Theta} = (\boldsymbol{\mathrm{R}}^{-1})^T \cdot \boldsymbol{C}_{\Phi}\cdot \boldsymbol{\mathrm{R}}^{-1}$ is defined in terms of the capacitance matrix~$[\boldsymbol{C}_{\Phi}]_{ij} = \partial^2 L(\boldsymbol{\Phi},\boldsymbol{\dot{\Phi}})/\partial\dot{\Phi}_i\partial\dot{\Phi}_j$, for~$i,j\in[0,N_J+1]$. Note that the~$\Sigma$ mode does not enter in the potential energy. 

After a Legendre transformation, we arrive at the circuit Hamiltonian
\begin{equation}
\begin{split}
H &= \boldsymbol{q}_{\Theta}^T\cdot \frac{\boldsymbol{C}_{\Theta}^{-1}}{2} \cdot \boldsymbol{q}_{\Theta} - \sum_{i=1}^{N_J} E_{J_i}\cos\theta_i \\
&- E_{J_{\mathrm{b}}} \cos\Bigg(\sum_{i=1}^{N_J}\theta_i + \varphi_{\mathrm{ext}}\Bigg),
\end{split}
\label{eq: superfluxo hamiltonian on difference-mode basis}
\end{equation}
where~$\boldsymbol{q}_{\Theta}\equiv\partial L(\boldsymbol{\Theta},\boldsymbol{\dot{\Theta}})/\partial\boldsymbol{\dot{\Theta}}=\boldsymbol{C}_{\Theta}\cdot\boldsymbol{\dot{\Theta}}$ is a vector of conjugate charge operators,~$\theta_i=\Theta_i/\varphi_0$ are phase operators and $\varphi_{\mathrm{ext}}=\Phi_{\mathrm{ext}}/\varphi_0$. In the presence of disorder in the circuit capacitances, the~$\sigma=\Sigma/\varphi_0$ mode couples slightly to the~$\theta_i$ modes via the respective conjugate charge operators. Here, we neglect this capacitive coupling under the assumption of small circuit-element disorder and a large-frequency $\sigma$ mode. The inverse capacitance matrix can thus be truncated to include only the~$\theta_i$ modes, i.e.~$\boldsymbol{C}_{\Theta}^{-1}\to\boldsymbol{C}^{-1}_{\Theta}[0:N_J-1,0:N_J-1]$, reducing~\cref{eq: superfluxo hamiltonian on difference-mode basis} to a Hamiltonian of~$N_J$ interacting degrees of freedom. Note that the resulting pairwise~$\theta_i$-$\theta_j$ capacitive coupling has all-to-all connectivity and exhibits no particular structure in the~$\theta_i$ basis.

\subsection{Accounting for charge dispersion}
\label{ss:Accounting for charge dispersion}

To model charge dispersion, we assume that each of the~$N_J+1$ circuit islands is coupled to a local fictitious voltage source~$V_i$ for~$i\in[0,N_J]$. The associated terms in the Lagrangian can generically be  written as~$\textstyle\sum_{i=0}^{N_J} ({C_{g_i}}/{2})(\dot{\Phi}_i-V_i)^2$, where~${C_{g_i}}$ is a gate capacitance for the~$i$th circuit node. Equivalently, this can be expressed as
\begin{equation}
L_g(\boldsymbol{\Phi},\boldsymbol{\dot{\Phi}}) = -\boldsymbol{\dot{\Phi}}^T\cdot\boldsymbol{C}_g\cdot \boldsymbol{V},
\label{eq: coupling lagrangian}
\end{equation}
where~$\boldsymbol{C}_g = \mathrm{diag}(C_{g_0}, C_{g_1}, \dots,C_{g_{N_J+1}})$ and~$\boldsymbol{V}=(V_0,V_1,\dots,V_{N_J+1})^T$. In addition to~\cref{eq: coupling lagrangian}, the capacitance matrix of the circuit is modified to account for the gate capacitances as~$\boldsymbol{{C}}_\Phi\to \boldsymbol{\widetilde{C}}_\Phi=\boldsymbol{{C}}_\Phi + \boldsymbol{C}_g$. 

Defining~$\boldsymbol{d}_{\Phi}=\boldsymbol{C}_g \cdot \boldsymbol{V}$, the conjugate charge operators are given by
\begin{equation}
\boldsymbol{q}_{\Theta} =\boldsymbol{\widetilde{C}}_{\Theta}\cdot\boldsymbol{\dot{\Theta}} - \boldsymbol{d}_{\Theta},
\label{eq: conjugate momenta charge dispersion}
\end{equation}
where~$\widetilde{\boldsymbol{C}}_{\Theta} = (\boldsymbol{\mathrm{R}}^{-1})^T \cdot \boldsymbol{\widetilde{C}}_{\Phi}\cdot \boldsymbol{\mathrm{R}}^{-1}$ and~$\boldsymbol{d}_{\Theta} = (\boldsymbol{\mathrm{R}}^{-1})^T\cdot \boldsymbol{d}_{\Phi}$. Note that due to charge conservation~$[\boldsymbol{d}_{\Theta}]_{N_J+1}=\textstyle\sum_{i=0}^{N_J}[\boldsymbol{d}_{\Phi}]_{i}/(N_J+1)$ is a constant of motion, and only~$N_J$ of the~$N_J+1$ offset charges are strictly independent. Using these expressions, the circuit Hamiltonian finally takes the form
\begin{equation}
\begin{split}
H &= (\boldsymbol{q}_{\Theta}+\boldsymbol{d}_{\Theta})^T\cdot \frac{\boldsymbol{\widetilde{C}}_{\Theta}^{-1}}{2} \cdot (\boldsymbol{q}_{\Theta}+\boldsymbol{d}_{\Theta}) \\
&- \sum_{i=1}^{N_J} E_{J_i}\cos\theta_i - E_{J_{\mathrm{b}}} \cos\Bigg(\sum_{i=1}^{N_J}\theta_i + \frac{\Phi_{\mathrm{ext}}}{\varphi_0}\Bigg).
\end{split}
\label{eq: superfluxo hamiltonian on difference-mode basis charge offset}
\end{equation}
Omitting the $\sigma$ mode and irrelevant constants, the above expression simplifies to 
\begin{equation}
\begin{split}
H &= \sum_{i=1}^{N_J}\Bigg[\frac{[\boldsymbol{\widetilde{C}}_{\Theta}^{-1}]_{ii}}{2}(q_i-q_{g_i})^2 -  E_{J_i}\cos\theta_i \Bigg] \\
&+ \sum_{j>i}^{N_J}[\boldsymbol{\widetilde{C}}_{\Theta}^{-1}]_{ij}q_iq_j - E_{J_{\mathrm{b}}} \cos\Bigg(\sum_{i=1}^{N_J}\theta_i + \varphi_{\mathrm{ext}}\Bigg),
\end{split}
\label{eq: superfluxo hamiltonian on difference-mode basis charge offset expanded}
\end{equation}
where~$q_{g_i} = [\boldsymbol{\widetilde{C}}_{\Theta}^{-1}\cdot \boldsymbol{d}_{\Theta}]_{i}/2[\boldsymbol{\widetilde{C}}_{\Theta}^{-1}]_{ii}$ for~$i\in[1,N_J]$ are effective offset charges in the~$\theta_i$ basis and~$q_i=[q_{\Theta}]_i$. This Hamiltonian is equivalent to~\cref{eq: fluxonium H}. Each of the bracketed terms in~\cref{eq: superfluxo hamiltonian on difference-mode basis charge offset expanded} define a site Hamiltonian ($H_{0_i}$ for the $i$th array junction), while the remaining terms correspond to both linear and nonlinear all-to-all interactions between the sites. Note that, in the main text, we have used the Cooper-pair-number operators $n_i=q_i/2e$ and the offset-charges $n_{g_i}=q_{g_i}/2e$, instead of $q_i$ and $q_{g_i}$, respectively. 

\section{Effective model for the fluxonium qubit}
\label{s:Effective models for the fluxonium qubit}

\subsection{Effective single-mode Hamiltonian}
\label{ss:Effective single-mode Hamiltonian}

We now derive an effective single-mode Hamiltonian for the fluxonium qubit that captures all circuit details. Because it is simple yet accurate, this model is used in the main text to assert the validity of the DMRG simulations in appropriate parameter ranges. 

To obtain this effective model, we first consider a change of coordinates in which adiabatically eliminating the circuit modes other than the superinductance mode~$\phi=\textstyle\sum_{i=1}^{N_J}\theta_i$ is simple. To find this appropriate change of coordinates, we reverse engineer the following Ansatz defining a new change of basis
\begin{equation}
\boldsymbol{\mathrm{R}}^{(1)} = 
\begin{pmatrix}
1-\sum_{k=1}^{N_J-1} a_k^{(1)} & 1+a_1^{(1)} & \cdots & 1+a_{N_J-1}^{(1)} & 0\\
-1 & 1 & 0 & \cdots & 0\\
\vdots & 0 & \ddots & \ddots & \vdots\\
-1 & \vdots & \ddots & 1 & 0\\
0 & 0 & 0 & 0 & 1\\
\end{pmatrix},
\label{eq: R prime matrix}
\end{equation}
where the constants~$\{a_k^{(1)}\}$ are defined by 
\begin{equation}
a_k^{(1)}=\frac{\sum_{i,j=0}^{N_J-1}(N_J[\boldsymbol{C}_{\Theta}]_{ik}\delta_{jk}-[\boldsymbol{C}_{\Theta}]_{ij})}{\sum_{i,j=0}^{N_J-1}[\boldsymbol{C}_{\Theta}]_{ij}},
\label{eq: ak constants}
\end{equation}
for~$k\in[1,N_J-1]$. Note that~\cref{eq: R prime matrix} acts as identity in the subspace of the~$\sigma$ mode and none of the~$\sigma$-mode components of the capacitance matrix~$\boldsymbol{C}_{\Theta}$ are included in~\cref{eq: ak constants}. The role of~$\boldsymbol{\mathrm{R}}^{(1)}$ is to  capacitively decouple a superinductance-like mode of the form 
\begin{equation}
\phi^{(1)}=\phi + \sum_{k=1}^{N_J-1} a_k^{(1)} (\theta_k-\theta_1),
\label{eq: superinductance-like mode}
\end{equation}
from all other circuit modes, while leaving the~$\sigma$ mode invariant. Indeed, the new capacitance matrix
\begin{equation}
\boldsymbol{C}_{X}^{(1)} = [(\boldsymbol{\mathrm{R}}^{(1)})^{-1}]^T \cdot \boldsymbol{C}_{X}^{(0)}\cdot (\boldsymbol{\mathrm{R}}^{(1)})^{-1},
\label{eq: CX1}
\end{equation}
with~$\boldsymbol{C}_{X}^{(0)}=\boldsymbol{C}_{\Theta}$ is block-diagonal in the absence of disorder. The first block has dimension~$1\times 1$ and corresponds to the~$\phi^{(1)}$ mode; the second block has dimension~$(N_J-1)\times (N_J-1)$ and involves all circuit modes except~$\phi^{(1)}$ and~$\sigma$; the last~$1\times 1$ block corresponds to the~$\sigma$ mode. By design, the first and second blocks of~\cref{eq: CX1} are exactly decoupled from each other, even in the presence of circuit-element disorder. In this case the first two blocks can be weakly coupled to the third block. Because the~$\sigma$ has a very high frequency for standard fluxonium circuit parameters, we neglect this coupling.

While the transformation~\cref{eq: R prime matrix} isolates the most relevant mode of the  circuit, we iterate recursively this transformation to decouple all remaining circuit modes in the capacitive interaction. Doing this will allow us to trace out such degrees of freedom later on. We proceed by defining an additional set of rotation matrices~$\{\boldsymbol{\mathrm{R}}^{(n)}\}$, for~$n\in[2,N_J-1]$, with the general form
\begin{widetext}
\begin{equation}
\boldsymbol{\mathrm{R}}^{(n)} = 
\begin{pmatrix}
1 & 0 & 0 & \cdots & \cdots & \cdots & \cdots & \cdots & \cdots & 0\\
0 & 1 & 0 & \cdots & \cdots & \cdots & \cdots & \cdots & \cdots & 0\\
\vdots & \ddots & \ddots & \ddots & \cdots & \cdots & \cdots & \cdots & \cdots & 0\\
\vdots & \vdots & 0 & 1 & 0 & \cdots & \cdots & \cdots & \cdots & 0\\
\vdots & \vdots & \vdots & 0 & 1-\sum_{k=n}^{N_J-1} a_k^{(n)} & 1+a_n^{(n)} & 1+a_{n+1}^{(n)} & \cdots & 1+a_{N_J-1}^{(n)} & 0\\
\vdots & \vdots & \vdots & \vdots & -1 & 1 & 0 & \cdots & 0 & \vdots\\
\vdots & \vdots & \vdots & \vdots & -1 & 0 & 1 & 0 & 0 & \vdots\\
\vdots & \vdots & \vdots & \vdots & \vdots & \vdots & \ddots & \ddots & \ddots & \vdots\\
\vdots & \vdots & \vdots & \vdots & -1 & 0 & \cdots & 0 & 1 & 0\\
0 & 0 & 0 & 0 & 0 & \cdots & \cdots & \cdots & 0 & 1
\end{pmatrix}.
\label{eq: R prime matrix n}
\end{equation}
\end{widetext}
Similarly to~$\boldsymbol{\mathrm{R}}^{(1)}$, the matrix~$\boldsymbol{\mathrm{R}}^{(n)}$ is composed by a~$n\times n$ identity block for the modes labeled by~$k<n$; a~$(N_J-n+1)\times (N_J-n+1)$ block for modes labeled by~$k\in[n,N_J-1]$; and a~$1\times 1$ block for the~$\sigma$ mode. The coefficients~$\{a_k^{(n)}\}$ are defined as
\begin{equation}
a_k^{(n)}=\frac{\sum_{i,j=n}^{N_J-1}\{(N_J-1+n)[\boldsymbol{C}_{X}^{(n-1)}]_{ik}\delta_{jk}-[\boldsymbol{C}_{X}^{(n-1)}]_{ij}\}}{\sum_{i,j=n}^{N_J-1}[\boldsymbol{C}_{X}^{(n-1)}]_{ij}},
\label{eq: ak constants general}
\end{equation}
which is a generalization of~\cref{eq: ak constants}. 

The transformations~$\boldsymbol{\mathrm{R}}^{(n<N_J-1)}$ are designed to each decouple a single mode, while~$\boldsymbol{\mathrm{R}}^{(N_J-1)}$ decouples the last two modes~$n=N_J-1$ and~$n=N_J$. Therefore, these~$N_J-1$ successive transformations exactly diagonalize the upper~$N_J\times N_J$ block of the capacitance matrix~$\boldsymbol{C}_{\Theta}$ that does not include the~$\sigma$ mode. We can then invert these transformations arriving at the expression
\begin{equation}
\theta_i=\frac{\phi^{(1)}}{N_J} + \sum_{n=2}^{N_J}v_{ni}\phi^{(n)},
\label{eq: phase across ith junction}
\end{equation}
where the coefficient~$v_{ni}$ quantifies how much the~$\phi^{(n)}$ mode couples to the~$i$th Josephson junction of the array. Using~\cref{eq: phase across ith junction} and the definition~$\phi=\sum_{i=1}^{N_J}\theta_i$ we moreover have
\begin{equation}
\phi=\phi^{(1)} + \sum_{n=2}^{N_J}\mathcal{V}_n\phi^{(n)},
\label{eq: phase across the black-sheep junction}
\end{equation}
where~$\mathcal{V}_n=\sum_{i=1}^{N_J}v_{ni}$. If~$C_0=0$, it follows that~$\mathcal{V}_n=0$ for~$n\in[2,N_J]$, and~$\phi^{(1)}\equiv \phi$ is the only mode that couples to the black-sheep junction. In other case, all modes are weakly coupled to the black-sheep junction, but this undesired coupling can be easily taken into account as we show in the following.

The relations~\cref{eq: phase across ith junction} and~\cref{eq: phase across the black-sheep junction} are now incorporated back to the potential energy of~\cref{eq: superfluxo hamiltonian on difference-mode basis}. In order to trace out the unwanted degrees of freedom, we write the operator~$\phi^{(n)}$ for~$n>1$ in terms of the harmonic-oscillator ladder operators as~$\phi^{(n)}=\sqrt{\pi z_n}(a_n + a_n^{\dagger})$. Here,~$z_n=\sqrt{L_n/C_n}/R_Q$ is the effective reduced impedance of the~$n$th mode, given  in terms of the effective inductance~$L_n$ and capacitance~$C_n$. While~$C_n$ can be readout directly from the block-diagonal capacitance matrix, the reduced inductance is determined by the product~$L_n^{-1}=\boldsymbol{X}_n^T\cdot(\boldsymbol{M}^{-1})^T\cdot\boldsymbol{L}^{-1}\cdot\boldsymbol{M}^{-1}\cdot\boldsymbol{X}_n$, where~$\bm{X}^{n}$ is the mode vector associated to~$\phi^{(n)}$ and~$\boldsymbol{M}=\textstyle(\prod_{n=1}^{N_J-1}\boldsymbol{R}^{(n)})^T\cdot\boldsymbol{R}$ is a matrix that reverses the multiple changes of basis. The trace can then be performed straightforwardly by noticing that 
\begin{equation}
e^{ix \phi^{(n)}} = e^{-\pi x^2 z_n /2} e^{i x \sqrt{\pi z_n}a_n^{\dagger}}e^{i x \sqrt{\pi z_n}a_n},
\label{eq: normal ordered exponential}
\end{equation}
and thus~$\mathrm{tr}_n[e^{ix \phi^{(n)}}\rho]=e^{-\pi x^2 z_n /2}$ where we assume that the~$n$th mode remains in its noninteracting vacuum state. Following to~\cref{eq: phase across ith junction} and~\cref{eq: phase across the black-sheep junction}, we approximate 
\begin{equation}
\begin{split}
\cos\theta_i&\simeq \mathrm{tr}_{n>1}[\cos\theta_i] \\
&\simeq x_i \cos[\phi^{(1)}/N_J],
\end{split}
\label{eq: normal ordered cosine}
\end{equation}
where~$x_i=\textstyle\prod_{n=2}^{N_J}e^{-\pi v_{ni}^2 z_n/2}$, and 
\begin{equation}
\begin{split}
\cos(\phi + \varphi_{\mathrm{ext}})&\simeq\mathrm{tr}_{n>1}[\cos(\phi + \varphi_{\mathrm{ext}})]\\
&\simeq x_{\mathrm{b}} \cos[\phi^{(1)} + \varphi_{\mathrm{ext}}],
\end{split}
\label{eq: normal ordered cosine black-sheep}
\end{equation}
with~$x_{\mathrm{b}} = \textstyle\prod_{n=2}^{N_J}e^{-\pi \mathcal{V}_{n}^2 z_n/2}$. In \cref{eq: normal ordered cosine,eq: normal ordered cosine black-sheep}, $\mathrm{tr}_{n>1}$ indicates a trace operation over all circuit modes $\phi^{(n)}$, except for $n=1$. Then, by renaming~$\phi^{(1)}\to\phi'$, we arrive at the effective single-mode Hamiltonian
\begin{equation}
\begin{split}
H&=4E_C n^{\prime 2}-\sum_{i=1}^{N_J} x_i E_{J_i}\cos(\phi'/N_J) \\
&- x_{\mathrm{b}} E_{J_{\mathrm{b}}}\cos(\phi'+\varphi_{\mathrm{ext}}),
\end{split}
\label{eq: effective fluxonium model}
\end{equation}
where~$E_C$ is taken to be the charging energy~$E_C=e^2/2[\boldsymbol{C}_{X}^{(1)}]_{00}$ of the~$\phi'$ mode and~$[\phi',n']=i$. Note that~\cref{eq: effective fluxonium model} is equivalent to~\cref{eq: effective fluxonium H} of the main text. Up to corrections of order~$N_J^{-3}$,~\cref{eq: effective fluxonium model} reduces to 
\begin{equation}
H=4E_C n^{\prime 2} + \frac{E_L}{2}\phi^{\prime 2} - E_J\cos(\phi'+\varphi_{\mathrm{ext}}),
\label{eq: effective fluxonium model simple}
\end{equation}
where~$E_L =  \textstyle\sum_{i=1}^{N_J}x_i E_{J_i}/N_J^2$ and~$E_J=x_{\mathrm{b}} E_{J_{\mathrm{b}}}$ are the effective inductive and Josephson-junction energies.~\cref{eq: effective fluxonium model simple} corresponds to the original fluxonium-qubit model of Ref.~\cite{manucharyan2009fluxonium}. Here, however, all energies entering~\cref{eq: effective fluxonium model simple} are specified by a precise function of the circuit-element parameters. 

\subsection{Qualitative regimes of the fluxonium qubit}
\label{ss:Qualitative regimes of the fluxonium qubit}

\begin{figure*}[t!]
\includegraphics[scale=1.]{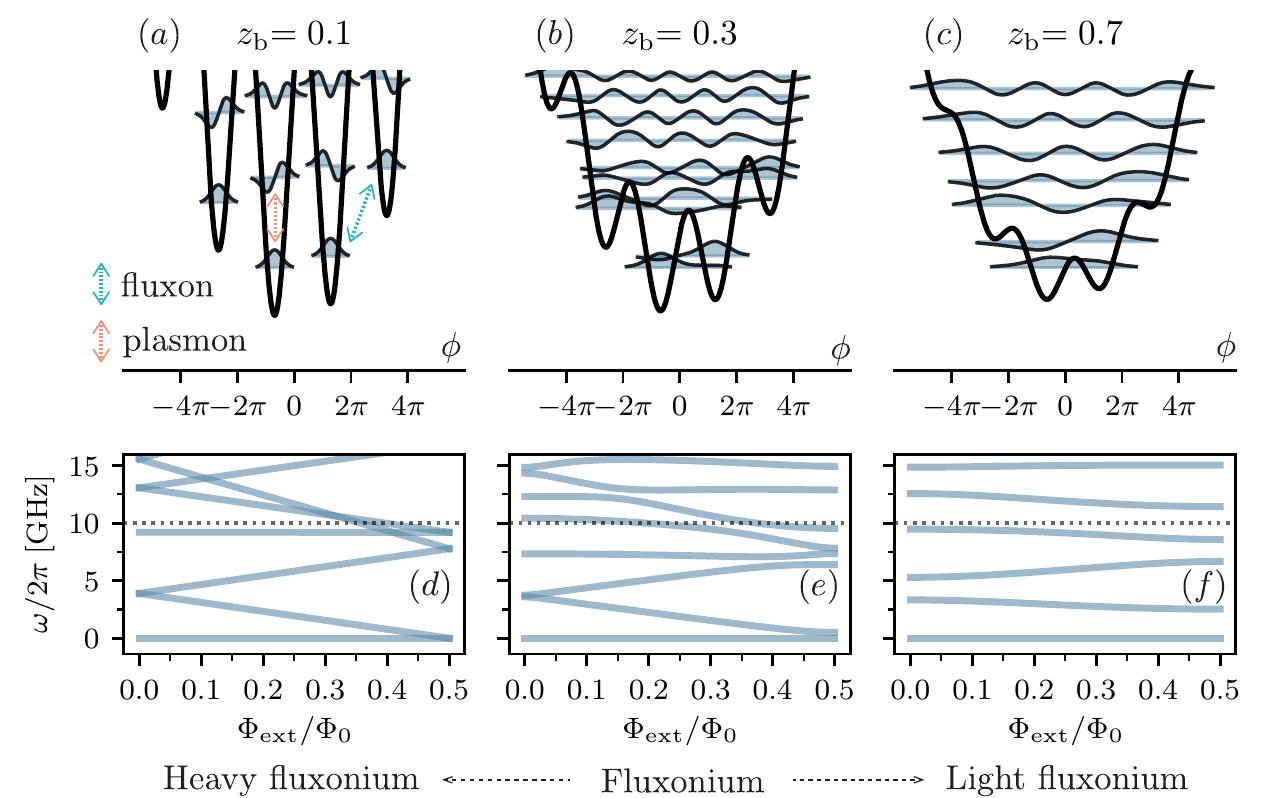}
\caption{\label{fig:wavefunction_details} Qualitative behavior of the eigenstates of the fluxonium qubit Hamiltonian~\cref{eq: effective fluxonium model simple}. ($a-c$) display the qubit wavefunctions (light-blue lines) within the effective potential (thick black line) for~$z_{\mathrm{b}}\in[0.1,0.3,0.7]$, respectively. ($d-f$) show the low-frequency spectrum as a function of~$\Phi_{\mathrm{ext}}$ (light-blue lines) for~$z_{\mathrm{b}}\in[0.1,0.3,0.7]$, respectively. Black dotted lines correspond to the bare black-sheep junction plasma frequency~$\omega_p^{\mathrm{b}}$. Additional parameters:~$\omega_p^{\mathrm{b}}/2\pi=10\,\mathrm{GHz}$ and~$E_L/h=0.2\,\mathrm{GHz}$.}
\end{figure*}

Despite the apparent simplicity of the effective fluxonium Hamiltonian~\cref{eq: effective fluxonium model simple}, its eigenstates can display a rich structure that depends on the parameter regime. For a systematic analysis, it is useful to redefine the parameters in~\cref{eq: effective fluxonium model simple} in terms of the effective black-sheep junction plasma frequency~$\omega_{p}^\mathrm{b}=\sqrt{8 E_J E_C}/\hbar$ and effective (reduced) impedance~$z_{\mathrm{b}}=\pi^{-1}\sqrt{2E_C/E_J}$. The potential energy of~\cref{eq: effective fluxonium model simple} has a quadratic component given by the inductive term~$E_L\phi^{2\prime}/2$ modulated by the cosine potential of the black-sheep junction and the external flux. Qualitatively,~$\hbar\omega_p^{\mathrm{b}}$ defines the characteristic energy of intra-well excitations within a given well defined by the Josephson potential, while~$z_{\mathrm{b}}$ is a measure of the tunneling amplitude between these wells. 

\Cref{fig:wavefunction_details} (a-c) shows the wavefunctions of the fluxonium qubit for different values of~$z_{\mathrm{b}}$, taking~$\omega_p^{\mathrm{b}}/2\pi=10\,\mathrm{GHz}$ and~$E_L/h=0.2\,\mathrm{GHz}$ constants and for~$\Phi_{\mathrm{ext}}/\Phi_0=0.35$. Panel ($a$) corresponds to the case of a small effective impedance with~$z_{\mathrm{b}}=0.1$, in which tunneling between states localized in different wells is exponentially suppressed~\cite{lin2018demonstration}. In this regime, the eigenstates of the fluxonium Hamiltonian are therefore localized within the deep potential wells of the potential-energy landscape. Excitations localized in a given potential well are approximately separated by the energy difference~$\hbar\omega_p^{\mathrm{b}}$. For this reason, a transition between two of such states is called plasmon (or intra-well) transition. On the other hand, a transition between two states that belong to different potential wells is called fluxon (or inter-well) transition. Since the relative positions between potential wells shift significantly with~$\Phi_{\mathrm{ext}}$, fluxon transitions are highly sensitive to the external flux. In contrast, plasmon transitions are only weakly flux-sensitive. Since the low-impedance limit requires the fluxonium mode~$\phi'$ to have a large effective capacitance (or ``mass''), this regime is referred to as `heavy-fluxonium' regime~\cite{earnest2018realization,lin2018demonstration,hazard2019nanowire}. 

\Cref{fig:wavefunction_details} ($b$) shows an intermediate value of~$z_{\mathrm{b}}=0.3$, where the energy barrier ($\propto E_J$) between the potential wells due to the black-sheep junction has been reduced with respect to panel ($a$). Moreover, the effective capacitive energy~$E_C$ has been increased, such that quantum tunneling between states localized in two neighboring potential wells is now non-negligible. This favors states that are delocalized across multiple potential wells and are the result of significant hybridization between plasmon and fluxon excitations. This intermediate regime for~$z_{\mathrm{b}}$ corresponds to the original fluxonium-qubit regime~\cite{manucharyan2009fluxonium,manucharyan2012superinductance}.

If the impedance of the black-sheep junction~$z_{\mathrm{b}}$ is increased further, the fluxonium wavefunctions can spread over many potential wells thanks to a lower~$E_J$ and a larger~$E_C$. This situation is illustrated in~\cref{fig:wavefunction_details} ($c$) where the distinction between plasmon and fluxon transitions is no longer useful and the spectrum is mostly determined by the harmonic part of~\cref{eq: effective fluxonium model simple}. The Josephson potential now leads to a weak flux sensitivity of the qubit transitions. Since the effective capacitance of the fluxonium mode needs to be lowered in order to make~$z_{\mathrm{b}}$ larger, this regime is known as the `light-fluxonium' regime~\cite{pechenezhskiy2019quantum}.

With the purpose of making the comparison above more precise, we now analyze qualitatively the energy spectrum of fluxonium devices from the heavy- to the light-fluxonium regimes. \cref{fig:wavefunction_details} ($d$) shows the result of the diagonalization of~\cref{eq: effective fluxonium model simple} (light-blue lines) for the parameters of~\cref{fig:wavefunction_details} ($a$). In this case, the low-frequency spectrum is highly sensitive to external flux, corresponding to a set of fluxon transitions. For a small~$z_{\mathrm{b}}$, the low-frequency spectrum around~$\Phi_{\mathrm{ext}}/\Phi_0=0.5$ can be modeled by the weak coupling of two ground states~$\{|m\rangle,|m+1\rangle\}$ with~$\langle\phi'|m\rangle \propto z_{\mathrm{b}}^{-1/4}\exp[-(\phi'-\phi_m)^2/4\pi z_{\mathrm{b}}]$ that are localized in two nearly degenerate potential wells with flux-dependent positions~$\{\phi_m\}$~\cite{catelani2011relaxation}. This model predicts a linear dispersion~$\propto 1/L$ of the first fluxon transition with the external flux, and a gap opening at~$\Phi_{\mathrm{ext}}/\Phi_0=0.5$ that is exponentially small in~$1/{z}_{\mathrm{b}}$~\cite{catelani2011relaxation}. In addition to the fluxon transitions,~\cref{fig:wavefunction_details} ($d$) reveals the first plasmon transition for the parameters in~\cref{fig:wavefunction_details} ($a$), corresponding to a flux-insensitive transition around~$\omega_p^{\mathrm{b}}/2\pi=10\,\mathrm{GHz}$. Since the nonlinearity of the black-sheep junction is small for low~$z_{\mathrm{b}}$, the plasmon transitions are only slightly shifted with respect to the bare plasma frequency~$\omega_p^{\mathrm{b}}$.

\Cref{fig:wavefunction_details} ($e$) shows the frequency spectrum corresponding to the parameters in~\cref{fig:wavefunction_details} ($b$). In this regime, the tunneling amplitude between different potential wells is stronger, leading to a larger hybridization gap between plasmon and fluxon transitions. However, for a moderate value of $z_{\mathrm{b}}$, the distinction between plasmon and fluxon transitions is still justified. As shown in~\cref{fig:wavefunction_details} ($f$), which corresponds to the spectrum associated to~\cref{fig:wavefunction_details} ($c$), this distinction is no longer convenient to interpret the case of a large black-sheep junction impedance. Indeed, as~$z_{\mathrm{b}}$ is made significantly larger, plasmon and fluxon transitions undergo a very strong hybridization. In this limit, the fluxonium eigenstates become insensitive to external magnetic flux, leading to a reduced flux dispersion of the qubit transition~\cite{koch2009charging,pechenezhskiy2019quantum}. 

\subsection{Exploration of various parameter regimes}
\label{ss:Exploration of various parameter regimes}

\begin{figure} [t!]
\includegraphics[scale=0.95]{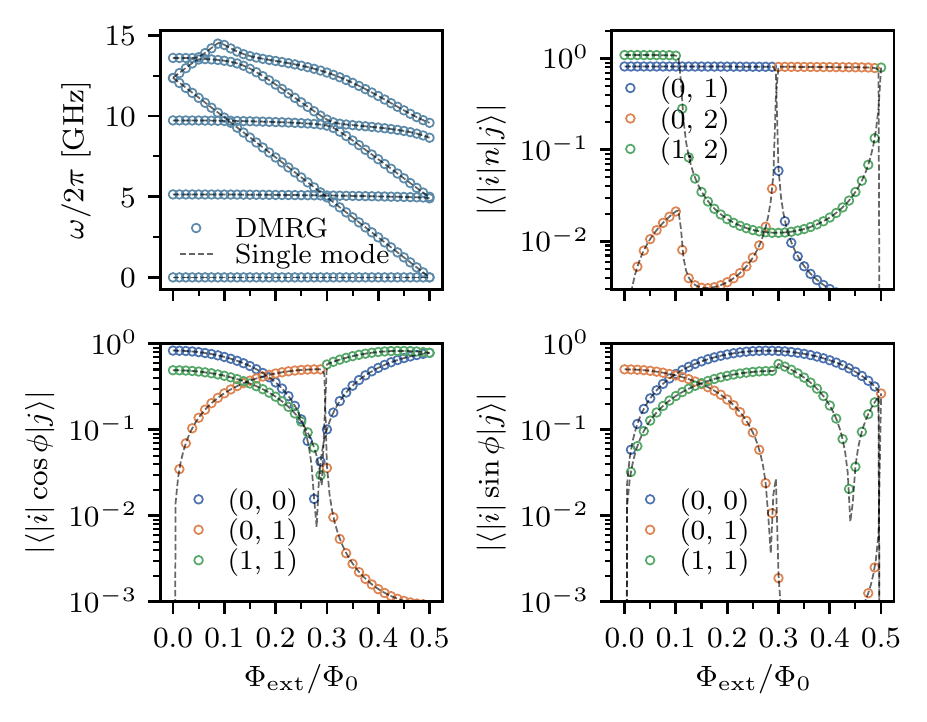}
\caption{\label{fig:fluxo_spectrum_sm} Comparison of results from~\cref{eq: superfluxo hamiltonian on difference-mode basis} (DMRG, circles) and~\cref{eq: effective fluxonium model} (Single-mode, black dashed lines) circuit Hamiltonians as a function of~$\Phi_{\mathrm{ext}}$. Top left panel: Energy spectrum. Top right panel: Matrix elements of the charge~$n$ operator for the superinductance mode. Bottom panels: Matrix elements of periodic functions of the phase~$\phi$ operator corresponding to the superinductance mode. DMRG parameters:~$N_J=180$,~$C_{J_{\mathrm{b}}}=40\,\mathrm{fF}$,~$E_{J_{\mathrm{b}}}/h=7.5\,\mathrm{GHz}$,~$C_J\simeq 32.9\,\mathrm{fF}$,~$L_J\simeq 1.23\,\mathrm{nH}$ (from~$\omega_p/2\pi=25\,\mathrm{GHz}$, and~$z = 0.03$) and~$C_0=0$. Single-mode model parameters:~$E_C/h\simeq 0.48\,\mathrm{GHz}$,~$E_L/h\simeq 1.27\,\mathrm{GHz}$ ($L\simeq 129.1\,\mathrm{nH}$) and~$E_J=E_{J_{\mathrm{b}}}$.}
\end{figure}

In this section, we provide further numerical evidence of the exceptional agreement between the DMRG simulations and the single-mode theory of~\cref{ss:Effective single-mode Hamiltonian}. For this purpose,~\cref{fig:fluxo_spectrum_sm} shows an extension of the results in the main text, including the spectrum of a fluxonium device with~$N_J=180$ array junctions and matrix elements of the phase and charge operators corresponding to the superinductance mode. As in the main body of the paper, the array junctions are modeled as multilevel systems including the first~$15$ eigenstates of the site Hamiltonian. The remarkable agreement between the DMRG simulation of the full model~\cref{eq: superfluxo hamiltonian on difference-mode basis} [symbols] and the effective single-mode Hamiltonian~\cref{eq: effective fluxonium model} [dashed lines] serves as a further validation of the DMRG results.

To demonstrate that the agreement between these two approaches extends to all parameter sets for which the array junctions behave as weakly anharmonic oscillators, we compare~\cref{eq: superfluxo hamiltonian on difference-mode basis} and~\cref{eq: effective fluxonium model} for various circuit design parameters. We also include the results obtained with an additional theory adapted from Ref.~\cite{hazard2019nanowire}, where the nonlinearity of the array junctions is not taken into account. More precisely, we employ a single-mode approximation of the multimode Hamiltonian of Ref.~\cite{hazard2019nanowire}, that we will refer to as `linear theory'. The objective of this additional comparison is to highlight the effect of the nonlinearity of the array junctions which, as shown below, renormalizes the effective superinductance. 

\begin{figure*} [t!]
\includegraphics[scale=1.]{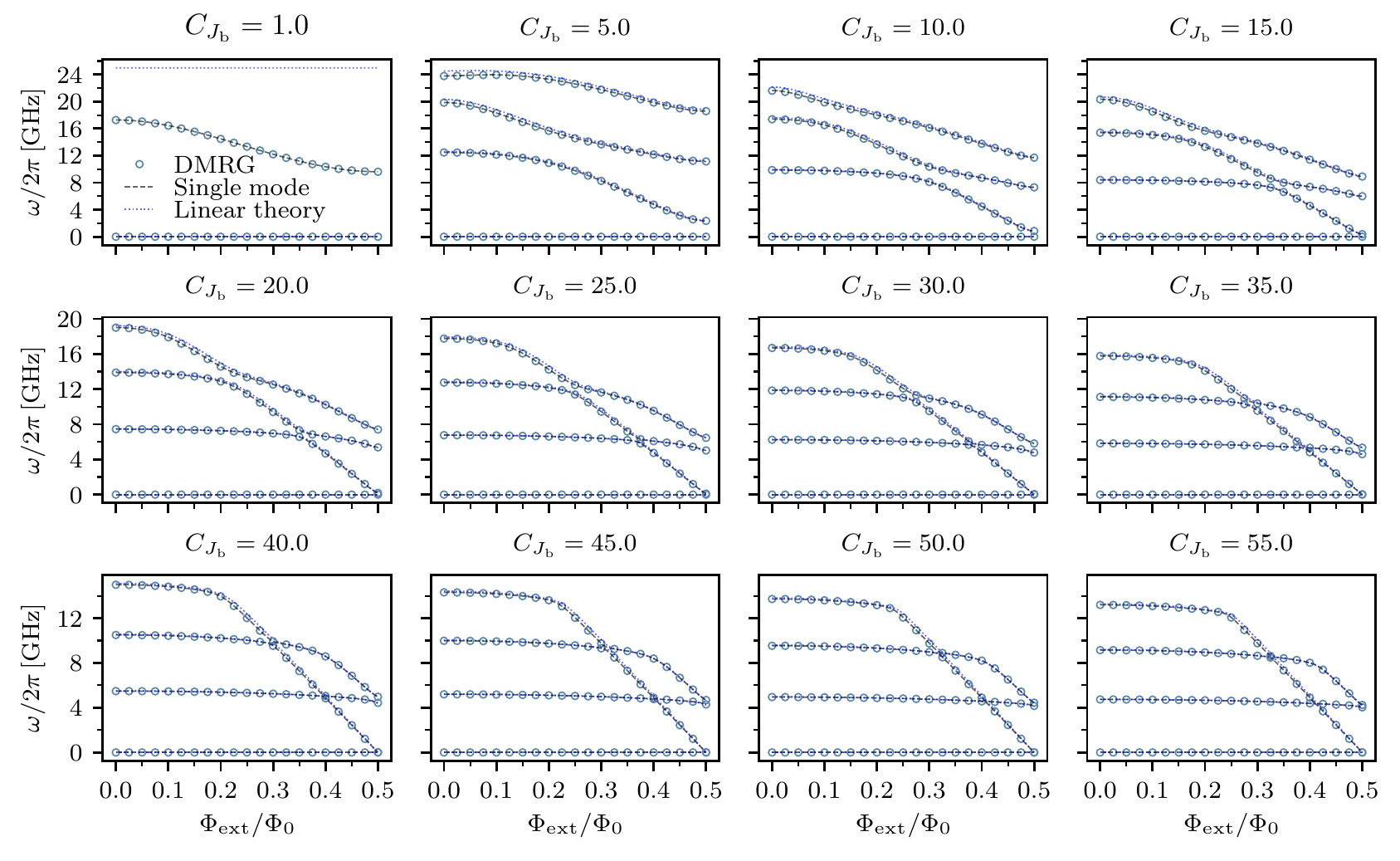}
\caption{\label{fig:sweep_cjbs} Comparison of results from~\cref{eq: superfluxo hamiltonian on difference-mode basis} (DMRG, circles),~\cref{eq: effective fluxonium model} (Single-mode, black dashed lines) and single-mode approximation based on the theory of Ref.~\cite{hazard2019nanowire} (Linear theory) for an~$80$-junction superinductance fluxonium device with a varying black-sheep capacitance in the range of~$C_{J_{\mathrm{b}}}\in[1,55]$ fF as a function of~$\Phi_{\mathrm{ext}}$. Additional parameters:~$E_{J_{\mathrm{b}}}/h=7.5\,\mathrm{GHz}$,~$C_J\simeq 32.9\,\mathrm{fF}$ and~$L_J\simeq 1.23\,\mathrm{nH}$ (from~$\omega_p/2\pi=25\,\mathrm{GHz}$ and~$z = 0.03$) and~$C_0=0$.}
\end{figure*}

\begin{figure*} [t!]
\includegraphics[scale=1.]{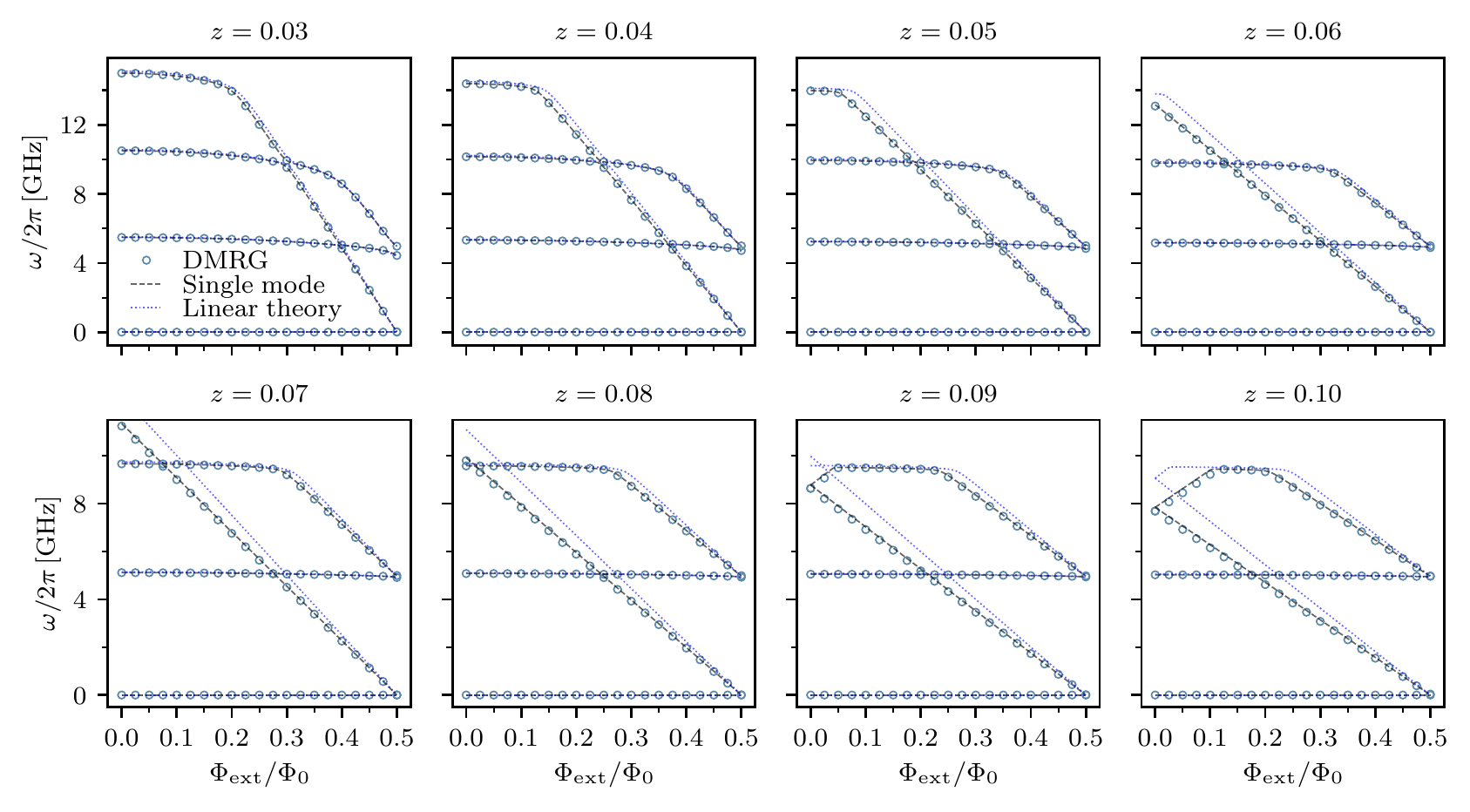}
\caption{\label{fig:sweep_z} Comparison of results from~\cref{eq: superfluxo hamiltonian on difference-mode basis} (DMRG, circles),~\cref{eq: effective fluxonium model} (Single-mode, black dashed lines) and single-mode approximation based on the theory of Ref.~\cite{hazard2019nanowire} (Linear theory) for an~$80$-junction superinductance fluxonium device with a varying array-junction reduced impedance in the range of~$z\in[0.03,0.10]$ as a function of~$\Phi_{\mathrm{ext}}$. Additional parameters:~$C_{J_{\mathrm{b}}}=40\,\mathrm{fF}$,~$E_{J_{\mathrm{b}}}/h=7.5\,\mathrm{GHz}$,~$\omega_p/2\pi=25\,\mathrm{GHz}$ and~$C_0=0$.}
\end{figure*}

In particular, we test circuit Hamiltonians for various black-sheep junction capacitances~(\cref{fig:sweep_cjbs}) and array-junction impedances~(\cref{fig:sweep_z}). The results of~\cref{fig:sweep_cjbs} demonstrate a very good agreement between the DMRG estimation~(symbols) and the single-mode Hamiltonian~\cref{eq: effective fluxonium model} [black dashed lines] from light- to heavy-fluxonium parameter sets. Blue dotted lines correspond to the predictions of the linear theory. Overall, the latter estimations are in good agreement with the DMRG and the single-mode-theory results, although we find appreciable deviations for some of the flux-sensitive transitions. As we discuss below, these deviations are explained by the effect of the array-junction nonlinearity.

\Cref{fig:sweep_z} shows a comparison between the results of DMRG, the single-mode theory and the linear theory for heavy-fluxonium parameter sets where the array-junction impedance is increased from $z=0.03$ to $z=0.10$. We observe that, in most cases, the single-mode theory of~\cref{ss:Effective single-mode Hamiltonian} provides an excellent estimation of the frequency of all fluxonium transitions determined by full DMRG. However, the predictive power of the effective single-mode theory weakens as~$z$ becomes larger (see~\cref{fig:sweep_z}~for~$z\gtrsim 0.08$). We attribute this discrepancy to the unfavorable scaling of the multimode coupling in~\cref{eq: superfluxo hamiltonian on difference-mode basis} with~$z$. This makes the approximation used to take the trace in~\cref{eq: normal ordered cosine} and~\cref{eq: normal ordered cosine black-sheep} not completely justified. Although further refinement of the theory of~\cref{ss:Effective single-mode Hamiltonian} might be possible, the breakdown of the noninteracting approximation defines a parameter regime where the DMRG estimations are in principle out of reach of a simple theory.

Furthermore,~\cref{fig:sweep_z} shows that the prediction of the linear theory of Ref.~\cite{hazard2019nanowire} [blue dotted lines] becomes increasingly inaccurate as $z$ increases. Indeed, since the renormalization of the effective superinductance scales exponentially with the array-junction impedance [see~\cref{eq: normal ordered cosine}], the frequency shifts of the qubit transitions due to the junction nonlinearity are more noticeable for larger~$z$. These frequency shifts are more clearly appreciated for the flux-sensitive (or fluxon) transitions in~\cref{fig:sweep_z}, which are highly sensitive to the effective superinductance value. This also explains the relatively small deviations encountered in~\cref{fig:sweep_cjbs} between the linear theory and DMRG for $z=0.03$.

Finally, we point out that we have also compared the result of the DMRG implementation to that of full exact diagonalization for fluxonium-like devices with a small number of junctions $N_J\in[2,6]$. We find excellent agreement between the DMRG and the exact-diagonalization implementations for all circuit parameters, strengthening the validity of our DMRG algorithm. These numerical tests provide solid evidence of a successful DMRG implementation of the full fluxonium Hamiltonian, thus complementing the results provided in the main text. 

\section{Multilevel pure-dephasing master equation for flux noise}
\label{s:Multilevel pure-dephasing master equation for flux noise}

In this section, we derive a master equation describing pure dephasing due to~$1/f$ flux noise in the fluxonium qubit. Assuming weak system-bath coupling, the master equation is obtained from the standard integro-differential equation 
\begin{equation}
\partial_t\rho(t) = -\frac{1}{\hbar^2}\int_{0}^{t}d\tau\,\mathrm{tr}_B[H_{\mathrm{int}}(t),[H_{\mathrm{int}}(t-\tau),\rho(t-\tau)\otimes\rho_B]],
\label{eq: master equation before Markov}
\end{equation}
where~$\rho(t)\otimes\rho_B$ is the system-bath density matrix, assumed to be separable at all times~\cite{breuer2002theory}. Assuming that the bath correlation functions are sharp around~$\tau= 0$,~$\rho(t-\tau)$ in~\cref{eq: master equation before Markov} can be approximated by~$\rho(t)$ with negligible error. This standard approximation conveniently leads to a Markovian master equation and allows us to extend the integral in~\cref{eq: master equation before Markov} to infinitely negative times. This last step is however not performed here in order to capture the Gaussian decay of the density matrix coherences in the presence of~$1/f$ noise.

The system-bath interaction Hamiltonian can be obtained from the fluxonium circuit Hamiltonian assuming that~$\Phi_{\mathrm{ext}}=\Phi_{\mathrm{ext}}^0 + \delta\Phi$, where~$\Phi_{\mathrm{ext}}^0$ is the applied flux bias and~$\delta\Phi$ represents fluctuations. To first order in~$\delta\Phi$, the interaction Hamiltonian can be written as~\cite{ithier2005}
\begin{equation}
H_{\mathrm{int}} = \partial_{\Phi_{\mathrm{ext}}} H|_{\Phi_{\mathrm{ext}}^0}\times\delta\Phi,
\label{eq: interaction H flux noise}
\end{equation}
where~$H$ is the Hamiltonian of the fluxonium qubit and the derivative with respect to the external flux is evaluated at~$\Phi_{\mathrm{ext}}=\Phi_{\mathrm{ext}}^0$. Expanding~\cref{eq: master equation before Markov} in the eigenbasis~$\{|\psi_k\rangle\}$ of the full circuit, we arrive at
\begin{widetext}
\begin{equation}
\partial_t\rho = -\frac{1}{\hbar^2}\sum_{\substack{k,k'\\l,l'}} \int_{0}^{t}d\tau\,\partial_{\Phi_{\mathrm{ext}}} H|_{\Phi_{\mathrm{ext}}^0}^{kk'}\partial_{\Phi_{\mathrm{ext}}} H|_{\Phi_{\mathrm{ext}}^0}^{ll'}e^{-i(\omega_{ll'}+\omega_{kk'})t+i\omega_{kk'}\tau}\mathrm{tr}_B[|\psi_l\rangle\langle\psi_{l'}|\delta\Phi(t),[|\psi_k\rangle\langle\psi_{k'}|\delta\Phi(t-\tau),\rho\otimes\rho_B]],
\label{eq: master equation eigenbasis}
\end{equation}
\end{widetext}
where we have introduced the matrix elements~$\partial_{\Phi_{\mathrm{ext}}} H|_{\Phi_{\mathrm{ext}}^0}^{kk'}=\langle\psi_k|\partial_{\Phi_{\mathrm{ext}}} H|_{\Phi_{\mathrm{ext}}^0}|\psi_{k'}\rangle$, and omitted the explicit time dependence of~$\rho(t)\to\rho$. 

Tracing out the bath degrees of freedom leads to the so-called Bloch-Redfield equation~\cite{breuer2002theory}. This equation has, however, a number of disadvantages that can potentially lead to unphysical dissipation results. Thus, for practical purposes, we use the rotating-wave approximation discarding terms for which~$\omega_{ll'}+\omega_{kk'}\neq 0$. As shown below, this approximation reduces~\cref{eq: master equation eigenbasis} to a Lindblad-form master equation. Assuming that the qubit has a set of nondegenerate energy transitions, this approximation is equivalent to the conditions~$l=k'$ and~$l'=k$ for~$\omega_{kk'}\neq 0$, and~$l=l'$ for~$\omega_{kk'}=0$. In this way,~\cref{eq: master equation eigenbasis} simplifies to 
\begin{widetext}
\begin{equation}
\begin{split}
\partial_t\rho =& -\frac{1}{\hbar^2}\sum_{k'>k}\int_{0}^{\infty}d\tau\, \partial_{\Phi_{\mathrm{ext}}} H|_{\Phi_{\mathrm{ext}}^0}^{kk'}\partial_{\Phi_{\mathrm{ext}}} H|_{\Phi_{\mathrm{ext}}^0}^{k'k}e^{i\omega_{kk'}\tau}\mathrm{tr}_B[|\psi_{k'}\rangle\langle\psi_{k}|\delta\Phi(t),[|\psi_k\rangle\langle\psi_{k'}|\delta\Phi(t-\tau),\rho\otimes\rho_B]]\\
&-\frac{1}{\hbar^2}\sum_{k'>k}\int_{0}^{\infty}d\tau\, \partial_{\Phi_{\mathrm{ext}}} H|_{\Phi_{\mathrm{ext}}^0}^{k'k}\partial_{\Phi_{\mathrm{ext}}} H|_{\Phi_{\mathrm{ext}}^0}^{kk'}e^{-i\omega_{kk'}\tau}\mathrm{tr}_B[|\psi_{k}\rangle\langle\psi_{k'}|\delta\Phi(t),[|\psi_{k'}\rangle\langle\psi_{k}|\delta\Phi(t-\tau),\rho\otimes\rho_B]]\\
&-\frac{1}{\hbar^2}\sum_{k,l}\int_{0}^{\infty}d\tau\, \partial_{\Phi_{\mathrm{ext}}} H|_{\Phi_{\mathrm{ext}}^0}^{kk}\partial_{\Phi_{\mathrm{ext}}} H|_{\Phi_{\mathrm{ext}}^0}^{ll}\mathrm{tr}_B[|\psi_{l}\rangle\langle\psi_{l}|\delta\Phi(t),[|\psi_k\rangle\langle\psi_{k}|\delta\Phi(t-\tau),\rho\otimes\rho_B]].
\end{split}
\label{eq: master equation eigenbasis rwa flux}
\end{equation}
\end{widetext}
We now assume that~$\delta\Phi(t)$ can be modeled as a (real) stationary random process. This assumption is motivated by physical models of bistable two-level-system defects that are known to produce noise of type~$1/f$~\cite{koch2007model,bialczak20071}. Furthermore, we make the usual assumption that the weight of the~$1/f$ noise spectral density is negligible at the qubit transition frequencies such that it does not significantly contribute to the device's~$T_1$ time. The pure-dephasing master equation is therefore derived from the third line of~\cref{eq: master equation eigenbasis rwa flux}, i.e.
\begin{equation}
\begin{split}
\partial_t\rho &=-\frac{1}{\hbar^2}\sum_{k,l}\int_{0}^{\infty}d\tau\, \partial_{\Phi_{\mathrm{ext}}} H|_{\Phi_{\mathrm{ext}}^0}^{kk}\partial_{\Phi_{\mathrm{ext}}} H|_{\Phi_{\mathrm{ext}}^0}^{ll}\\
&\times\mathrm{tr}_B[|\psi_{l}\rangle\langle\psi_{l}|\delta\Phi(t),[|\psi_k\rangle\langle\psi_{k}|\delta\Phi(t-\tau),\rho\otimes\rho_B]].
\end{split}
\label{eq:pure dephasing master equation eigenbasis rwa}
\end{equation}
Next, we introduce the noise spectral density $S^{1/f}_\Phi[\omega]$ for $1/f$~flux noise by the definition \cite{devoret1995quantum} 
\begin{equation}
\mathrm{tr}_B[\rho_B\delta\Phi(t)\delta\Phi(t')]=\frac{1}{2\pi}\int_{-\infty}^{\infty}d\omega\,S^{1/f}_\Phi[\omega]e^{-i\omega(t-t')},
\label{eq:noise spectral density definition}
\end{equation}
and assume the general form
\begin{equation}
S^{1/f}_\Phi(\omega) = \frac{A_\Phi^2}{|\omega|/2\pi},
\label{eq: spectral density 1/f noise}
\end{equation}
where~$A_\Phi$ is the~$1/f$~flux-noise amplitude, typically reported to be in the range~$1-10\,\mu\Phi_0$~\cite{koch2007charge}. It must be stressed that \cref{eq: spectral density 1/f noise} is an approximation to the spectral densities measured in the laboratory, which can scale as $|\omega|^{-\mu}$ with $\mu\in[0.6,1.3]$ \cite{koch2007charge,slichter2012measurement}.

We proceed further by exploiting a simple mathematical fact. Using \cref{eq:noise spectral density definition} and \cref{eq: spectral density 1/f noise}, we find that 
\begin{equation}
\int_0^td\tau\,\mathrm{tr}_B[\rho_B\delta\Phi(t)\delta\Phi(t')]=\lim_{\omega_{\mathrm{ir}}\to 0}-2A_\Phi^2\int_{0}^{t} d\tau\,\mathrm{Ci}(\omega_{\mathrm{ir}}\tau),
\label{eq:integral of correlation function}
\end{equation}
where~$\mathrm{Ci}(y) = -\int_y^{\infty}dx\,x^{-1}\cos x$ is the cosine integral. Here,~$\omega_{\mathrm{ir}}$ is an infrared frequency cutoff in the order of $2\pi \times 1\,\mathrm{Hz}$, introduced to regularize the cosine integral and motivated by physical reasons~\cite{groszkowski2018coherence}. Since the time~$t$ over which we are interested in calculating the time evolution of the density matrix is small compared to the time scale set by~$\omega_{\mathrm{ir}}^{-1}$, we make use of the series expansion
\begin{equation}
\mathrm{Ci}(w) = \gamma + \log(y) + \sum_{k=1}^{\infty}\frac{(-y^2)^k}{2k(2k)!},
\label{eq: cosine integral series}
\end{equation}
where~$\gamma\simeq 0.58$ is the Euler's constant, approximating
\begin{equation}
\int_0^td\tau\,\mathrm{tr}_B[\rho_B\delta\Phi(t)\delta\Phi(t')]\simeq 2 A_\Phi^2\,t\,[(1-\gamma)-\log(\omega_{\mathrm{ir}}t)].
\label{eq:integral of correlation function approximated}
\end{equation}

Expanding the double commutators in~\cref{eq:pure dephasing master equation eigenbasis rwa} and making use of~\cref{eq:integral of correlation function approximated}, we arrive at a pure-dephasing master equation of the form
\begin{equation}
\begin{split}
\partial_t\rho &= \sum_{k} \Gamma_{\varphi}^{kk}\,\mathcal{D}[\sigma_{kk},\sigma_{kk}]\,\rho \\
& + \sum_{k>l} \Gamma_{\varphi}^{kl} \Big(\mathcal{D}[\sigma_{kk},\sigma_{ll}]+\mathcal{D}[\sigma_{ll},\sigma_{kk}]\Big)\rho,
\end{split}
\label{eq: pure dephasing lindbladian}
\end{equation}
where $\Gamma_{\varphi}^{kl}$ are time-dependent pure-dephasing rates given by
\begin{equation}
\Gamma_{\varphi}^{kl} = \partial_{\Phi_{\mathrm{ext}}} H|_{\Phi_{\mathrm{ext}}^0}^{kk}\partial_{\Phi_{\mathrm{ext}}} H|_{\Phi_{\mathrm{ext}}^0}^{ll}\times 4A_{\Phi}^2\,t\,[(1-\gamma) - \log(\omega_\mathrm{ir} t)],
\label{eq: pure dephasing rates classical}
\end{equation}
$\sigma_{kl}=|\psi_k\rangle\langle\psi_l|$, and~$\mathcal{D}[x,y]\,\rho = x\rho y^{\dagger}-\{y^{\dagger}x,\rho\}/2$ is a generalized dissipator superoperator. Equivalently,~\cref{eq: pure dephasing lindbladian} can be recast in the more familiar form
\begin{equation}
\begin{split}
\partial_t\rho &= \sum_{k} \Gamma_{\varphi}^{kk}\,\mathcal{D}[\sigma_{kk}]\,\rho \\
&+ \sum_{k>l} \Gamma_{\varphi}^{kl}\,\Big(\mathcal{D}[\sigma_{kk}+\sigma_{ll}]-\mathcal{D}[\sigma_{kk}]-\mathcal{D}[\sigma_{ll}]\Big)\rho,
\end{split}
\label{eq: pure dephasing lindbladian form classical}
\end{equation}
where~$\mathcal{D}[x]\,\rho = x\rho x^{\dagger}-\{x^{\dagger}x,\rho\}/2$ is the standard dissipator superoperator. By projecting~\cref{eq: pure dephasing lindbladian form classical}, one has
\begin{equation}
\langle \psi_k|\partial_t\rho|\psi_l\rangle = -\frac{1}{2}\Big[\Gamma_{\varphi}^{kk}+\Gamma_{\varphi}^{ll}-2\Gamma_{\varphi}^{kl}\Big]\langle \psi_k|\rho|\psi_l\rangle,
\label{eq: pure dephasing lindbladian form kl}
\end{equation}
where 
\begin{equation}
\Big[\Gamma_{\varphi}^{kk}+\Gamma_{\varphi}^{ll}-2\Gamma_{\varphi}^{kl}\Big]\propto[\partial_{\Phi_{\mathrm{ext}}} (\hbar\omega_{kl})|_{\Phi_{\mathrm{ext}}^0}]^2.
\label{eq: proportionality with dispersion}
\end{equation}
Thus, we verify that the decay of the coherences of the density matrix is proportional to the flux dispersion of the~$k\leftrightarrow l$ qubit transition, as expected for first-order dephasing processes. Since second-order corrections to the pure-dephasing rate at sweet spots are of order~$A_\Phi^4$ and vanishing small, most devices are~$T_1$-limited at such operating points. Now, in order to produce an estimate of the pure-dephasing coherence time due to~$1/f$ flux noise, we simply integrate~\cref{eq: pure dephasing lindbladian form kl}, arriving at the expression
\begin{widetext}
\begin{equation}
\rho_{kl}(t)=\rho_{kl}(0)\exp\Big\{-A_\Phi^2(\partial_{\Phi_{\mathrm{ext}}}\omega_{kl}|_{\Phi_{\mathrm{ext}}^0})^2\,t^2\,\Big[\Big(\frac{3}{2}-\gamma\Big)-\log(\omega_{\mathrm{ir}}t)\Big]\Big\}.
\label{eq: pure dephasing lindbladian form kl integrated out}
\end{equation}
\end{widetext}
We note that expressions similar to~\cref{eq: pure dephasing lindbladian form kl integrated out} have been derived previously in the literature~\cite{koch2007charge,groszkowski2018coherence}. However, these expressions do not include the correction~$(\frac{3}{2}-\gamma)$ within brackets in~\cref{eq: pure dephasing lindbladian form kl integrated out}. Finally, we define the coherence time~$T_{\varphi}$ as the solution of the implicit equation~$\rho_{01}(T_{\varphi})/\rho_{01}(0)=1/e$. The solution of this equation has been used in~\cref{fig:vlad_device} to produce an estimation of the pure-dephasing coherence times due to flux noise. 

\bibliography{library}

\end{document}